\documentclass[a4paper,11pt]{article}
\pdfoutput=1 
\usepackage{jheppub}
\usepackage[T1]{fontenc} 
\usepackage{color}
\usepackage{fancybox}
\usepackage{amsmath}
\usepackage{amsfonts}
\usepackage{slashed}
\usepackage{amssymb}
\usepackage{indentfirst}
\usepackage{stmaryrd}
\usepackage{subcaption}
\usepackage{graphicx}
\usepackage{cleveref}
\usepackage[toc]{appendix}
\usepackage[font={small,it}]{caption}
\usepackage[justification=centering]{caption}

\allowdisplaybreaks

\newcommand{\be}{\begin{equation}}
\newcommand{\bea}{\begin{eqnarray}}

\newcommand{\ee}{\end{equation}}
\newcommand{\eea}{\end{eqnarray}}

\title{\boldmath On the equivalence of the 11D pure spinor and Brink-Schwarz-like superparticle cohomologies}

\author{Max Guillen$^{*}$}

\affiliation{$^{*}$Instituto de F\'{i}sica Te\'{o}rica, UNESP-Universidade Estadual Paulista\\ R. Dr. Bento T. Ferraz 271, Bl. II, S\~{a}o Paulo 01140-070, SP, Brazil}

\emailAdd{luismax@ift.unesp.br}

\abstract{The $D=11$ pure spinor formulation of the superparticle allows a simple realization of covariant quantization, unlike the $D=11$ Brink-Schwarz-like superparticle. We explicitly show the equivalence between the cohomologies of these two models in the context of two different group decompositions: $SO(10,1) \rightarrow SO(1,1)\times SO(9)$ and $SO(10,1) \rightarrow SO(3,1)\times SO(7)$. We also carry out a light-cone analysis of the pure spinor cohomology, and show that it correctly reproduces the $SO(9)$ equations of motion for $D=11$ linearized supergravity.
}

\keywords{Supergravity, Superparticle, Pure spinors.}

\begin{document} 
\hfill{}
\maketitle

\section{Introduction}

It is well known that the $D=10$ Brink-Schwarz formulation of the superparticle possesses first- and second-class constraints which cannot be separated out in a manifestly covariant way. If the physical spectrum is our main concern, we can always go to the light-cone gauge and follow Dirac's prescription to show that the physical spectrum consists of an $SO(8)$ vector and spinor, which satisfy the $D=10$ linearized Super Yang-Mills equations of motion \citep{Brink:1981nb}. However, light-cone gauge breaks the manifest covariance of the theory.
\vspace{2mm}

It is interesting and useful to look for covariant descriptions which manifestly preserve as many symmetries as possible. One candidate that addresses this point is the pure spinor version of the $D=10$ Brink-Schwarz superparticle, known as the \emph{D=10 pure spinor superparticle} \cite{Berkovits:2001rb,Bedoya:2009np}. This description preserves supersymmetry and Lorentz symmetry in a manifestly covariant way. The spectrum is defined as the cohomology of the BRST operator defined by $Q = \lambda^{\mu} d_{\mu}$, where $\lambda^{\mu}$ is a $D=10$ pure spinor
and the $d_{\mu}$ are the fermionic constraints of the $D=10$ Brink-Schwarz superparticle. There are two ways to see that the pure spinor formulation indeed describes $D=10$ linearized Super Yang-Mills. The first one is by looking at the $Q$-cohomology of the $D=10$ pure spinor superparticle and realizing that the elements in this cohomology describe the BV version of $D=10$ (abelian) Super Yang-Mills \cite{Berkovits:2001rb}. The second one is by showing that the cohomologies corresponding to the $D=10$ Brink-Schwarz superparticle and the $D=10$ pure spinor superparticle are identical \cite{Berkovits:2002zk}. 

\vspace{2mm}
As explained in \cite{Berkovits:2001rb,Berkovits:2002zk,Berkovits:2002uc,Bedoya:2009np}, the $D=10$ SYM physical fields are found in the ghost-number 1 vertex operator $V = \lambda^{\mu}A_{\mu}$, after imposing on it the pure spinor physical state condition. The light-cone analysis of this cohomology reproduces the $SO(8)$ superfield $A_{a}$ satisfying the SYM equations of motion in $D=8$ superspace \cite{Berkovits:2014bra}.

\vspace{2mm}
In $D=11$ the story is similar. The $D=11$ Brink-Schwarz-like superparticle \cite{Green:1999by} possesses first-class and second-class constraints which do not allow a manifestly covariant quantization of the theory. However, it is possible to quantize the theory in the light-cone gauge and it can be shown that the spectrum is described by an $SO(9)$ traceless symmetric tensor, an $SO(9)$ $\Gamma$-traceless vector-spinor and an $SO(9)$ 3-form which describe $D=11$ linearized Supergravity. As before, this theory is no longer manifestly Lorentz covariant.

\vspace{2mm}
As in the $D=10$ case, Berkovits formulated the so-called \emph{$D=11$ pure spinor superparticle} \cite{Berkovits:2002uc}. The physical states of this pure spinor version are defined as elements in the cohomology of the BRST operator $Q = \Lambda^{\alpha}D_{\alpha}$, where $\Lambda^{\alpha}$ is a $D=11$ pure spinor and $D_{\alpha}$ are the fermionic constraints of the $D=11$ Brink-Schwarz-like superparticle. The elements of this $Q$-cohomology describe the BV version of $D=11$ linearized supergravity \cite{Berkovits:2002uc}. Unlike the $D=10$ case there is not explicit proof of the equivalence between the cohomologies of the $D=11$ Brink-Schwarz-like superparticle and the $D=11$ pure spinor superparticle\footnote{There is a brief discussion of this point in \cite{Anguelova:2004pg}, which suggests following the same ideas developed in the $D=10$ case. We will elaborate on the ideas mentioned there, and give another way to parametrize $D=11$ pure spinors.}. In this work we will demonstrate the equivalence of these two cohomologies by using two different group decompositions\footnote{In \cite{Bandos:2007mi,Bandos:2007wm} I. Bandos relates these two models by using the Lorentz harmonics approach. We will address the problem in a different way, by focusing on the $D=11$ light-cone Brink-Schwarz-like superparticle.}.

\vspace{2mm}
As explained in \cite{Berkovits:2002uc}, the $D=11$ supergravity physical fields are found in the ghost number 3 vertex operator $V = \Lambda^{\alpha}\Lambda^{\beta}\Lambda^{\delta}C_{\alpha\beta\delta}$, after imposing the pure spinor physical state condition. The light-cone analysis of this cohomology will be described by the $SO(9)$ superfields $g_{jk}$, $\tilde{\psi}^{j}_{A}$, $C_{jkl}$, which satisfy a set of equations of motion in $D=9$ superspace that match the linearized supergravity light-cone equations of motion \cite{Green:1999by}.

\vspace{2mm}
The paper is organized as follows: In section 2 we review the $D=11$ Brink-Schwarz-like superparticle. In section 3 we present the $D=11$ pure spinor superparticle and show the equivalence between the cohomologies of this theory and the previous one by decomposing $D=11$ objects into their $SO(1,1)\times SO(9)$ and $SO(3,1)\times SO(7)$ components. In section 4 we study the light-cone pure spinor cohomology and show that it is described by the usual $SO(9)$ irreducible representations that describe $D=11$ supergravity and satisfy linearized equations of motion in $D=9$ superspace.

\section{Review of the $D=11$ Brink-Schwarz-like superparticle}
The $D=11$ Brink-Schwarz-like superparticle is defined by the action \cite{Green:1999by,Berkovits:2002uc}:
\begin{equation}\label{eq60}
S = \int d\tau(P^{m}\Pi_{m} + eP^{m}P_{m})
\end{equation}
where $\Pi_{m} = \dot{X}_{m} - \dot{\Theta}^{\alpha}(\Gamma_{m})_{\alpha\beta}\Theta^{\beta}$, and $\Theta^{\alpha}$ is a Majorana spinor. Let us now fix conventions. We will denote $SO(10,1)$ vector indices by $m, n, p, \ldots$, and spinor indices by $\alpha,\beta, \ldots$ ($m = 0, \ldots, 10$ and $\alpha = 1, \ldots, 32$). The $D=11$ gamma matrices $\Gamma^{m}$ are $32\times 32$ symmetric matrices which satisfy $\Gamma^{m}_{\alpha\beta}\Gamma^{n\,\beta\gamma} + \Gamma^{n}_{\alpha\beta}\Gamma^{m\,\beta\gamma} = 2\eta^{mn}\delta_{\alpha}^{\gamma}$ and $\eta_{mn}\Gamma^{m}_{(\alpha\beta}\Gamma^{np}_{\gamma\delta)} = 0$. In contrast to the $D=10$ case, in $D=11$ there exists an antisymmetric metric tensor $C_{\alpha\beta}$ (and its inverse $(C^{-1})^{\alpha\beta}$) which will allow us to lower (and raise) indices (for instance $\Gamma^{m\,\alpha\beta} = C^{\alpha\delta}\Gamma^{m\,\beta}_{\delta}$, etc). We also note that any $D=11$ antisymmetric bispinor can be decomposed into a scalar, three-form, and four form as $f^{[\alpha\beta]} = C^{\alpha\beta}f + (\Gamma_{mnp})^{\alpha\beta}f^{mnp} + (\Gamma_{mnpq})^{\alpha\beta}f^{mnpq}$, and that any $D=11$ symmetric bispinor can be written in terms of a one-form, two-form and five-form as $g^{(\alpha\beta)} = \Gamma^{\alpha\beta}_{m}g^{m} + (\Gamma_{mn})^{\alpha\beta}g^{mn} + (\Gamma_{mnpqr})^{\alpha\beta}g^{mnpqr}$.\\

The action \eqref{eq60} is invariant under reparametrizations, SUSY transformations and $\kappa$-transformations which are defined by the following equations:
\begin{eqnarray*}
\mbox{Reparametrizations} &\rightarrow & d\tau\ensuremath{'} = \frac{d\tau\ensuremath{'}}{d\tau}d\tau \hspace{2mm} \mbox{,} \hspace{2mm} e\ensuremath{'}(\tau)=\frac{d\tau}{d\tau\ensuremath{'}}d\tau\\
\mbox{SUSY transformations} &\rightarrow& \delta\Theta^{\alpha} = \epsilon^{\alpha} \hspace{2mm} \mbox{,} \hspace{2mm} \delta X^{m} = \Theta^{\alpha}\Gamma^{m}_{\alpha\beta}\epsilon^{\beta} \hspace{2mm} \mbox{,} \hspace{2mm} \delta P_{m} = \delta e = 0\\
\mbox{$\kappa$ (local) transformations} &\rightarrow& \delta\Theta^{\alpha} = iP^{m}\Gamma^{\alpha\beta}_{m}\kappa_{\beta} \hspace{2mm} \mbox{,} \hspace{2mm} \delta X^{m} = -\Theta^{\alpha}\Gamma^{m}_{\alpha\beta}\delta\Theta^{\beta} \hspace{2mm} \mbox{,} \hspace{2mm} \delta P_{m} = 0 ,\\
&& \delta e = 2i\dot{\Theta}^{\beta}\kappa_{\beta}
\end{eqnarray*}
The conjugate momentum to $\Theta^{\alpha}$ is
\begin{equation}
P_{\alpha} = \frac{\partial L}{\partial \dot{\Theta}^{\alpha}} = -\Gamma^{m}_{\alpha\beta}\Theta^{\beta}P_{m}
\end{equation}
Therefore, this system possesses constraints,
\begin{equation}\label{eq501}
D_{\alpha} = P_{\alpha} + \Gamma^{m}_{\alpha\beta}\Theta^{\beta}P_{m}
\end{equation}
and considering that 
 $\{\Theta^{\alpha}, P_{\beta}\}_{P.B} = i\delta^{\alpha}_{\beta}$, we get the constraint algebra
\begin{equation}\label{eq61}
\{D_{\alpha}, D_{\beta}\} = 2i(\Gamma^{m})_{\alpha\beta}P_{m},
\end{equation}
where $\{\cdot ,\cdot \}$ denotes a Poisson bracket. One can show that $K^{\alpha} = P^{m}\Gamma_{m}^{\alpha\beta}D_{\beta}$ are the first-class constraints that generate the $\kappa$-symmetry.
From \eqref{eq61}, we realize that we have 16 first-class constraints and 16 second-class constraints, and there is no simple way to covariantly separate them out. However, the physical spectrum can be easily found by using the semi light-cone gauge, which is defined by:
\begin{eqnarray}
X^{+} = \frac{1}{\sqrt{2}}(X^{0} + X^{9}) \hspace{4mm}&,&\hspace{4mm} \Gamma^{+} = \frac{1}{\sqrt{2}}(\Gamma^{0} + \Gamma^{9})\\
X^{-} = \frac{1}{\sqrt{2}}(X^{0} - X^{9}) \hspace{4mm}&,&\hspace{4mm} \Gamma^{-} = \frac{1}{\sqrt{2}}(\Gamma^{0} - \Gamma^{9})
\end{eqnarray}
In these light-cone coordinates one can use the $\kappa$-transformation to choose a gauge where $(\Gamma^{+}\Theta)_{\alpha} = 0$\footnote{An easy way to see this is to choose a frame where $P^{m}=(P,0,\ldots,P,0)$. The $\kappa$-transformation takes the form $\delta\Theta^{\alpha} = -iP^{+}\Gamma^{-\,\alpha\beta}\kappa_{\beta}$, and thus it follows immediately that $(\Gamma^{+}\Theta)_{\alpha} = 0$.}. With this choice we can rewrite the action as follows
\begin{equation}\label{eq72}
S = \int d\tau[P^{m}\dot{X}_{m} - \frac{i}{2}S^{A}S_{A} + e P^{m}P_{m}]
\end{equation}

where $S_{A}$ is an $SO(9)$ Majorana spinor, which can be written in terms of $SO(9)$ component of $\Theta^{\alpha}$. The conjugate momentum to $S^{A}$ is:
\begin{equation}
p_{A} = \frac{\partial L}{\partial \dot{S}^{A}} = -\frac{i}{2}S_{A}
\end{equation}
So, the constraints for this gauge-fixed system are:
\begin{equation}
\tilde{D}_{A} = p_{A} + \frac{i}{2}S_{A}
\end{equation}
Considering that $\{S_{A}, p_{B}\} = -i\delta_{AB}$, we obtain
\begin{eqnarray}
\{\tilde{D}_{A}, \tilde{D}_{B}\} &=& \delta_{AB}
\end{eqnarray}
Hence, the constraint matrix is $C_{AB} = \delta_{AB}$, and its corresponding inverse is $(C^{-1})^{AB} = \delta^{AB}$. This allows us to compute the following Dirac Bracket:
\begin{eqnarray}
\{S_{A}, S_{B}\}_{D} &=& \{S_{A}, S_{B}\}_{P} - \sum_{E,F}\{S_{A}, \tilde{D}_{E}\}_{P}(C^{-1})^{EF}\{\tilde{D}_{F}, S_{B}\}_{P}\nonumber\\
&=& 0 - \sum_{E,F}(-i\delta_{AE})(\delta^{EF})(-i\delta_{FB})\nonumber\\
&=& \delta_{AB} \label{eq71}
\end{eqnarray}
As is well known, the representation of the algebra \eqref{eq71} defines the space of physical states. These states will be denoted $\vert IJ\rangle$, $\vert BI\rangle$ and $\vert LMN\rangle$, where we represent $SO(9)$ vector indices by $I, J, K, L, \ldots$, and spinor indices by $A, B, C, D, \ldots$. These states correspond to an $SO(9)$ traceless symmetric tensor, an $SO(9)$ $\Gamma$-traceless vectorspinor and an $SO(9)$ 3-form, which, together, form the field content of $D=11$ SUGRA. The action of the operators $S_{A}$ on the physical states is defined by
\begin{eqnarray}
S_{A}\vert IJ\rangle &=& \Gamma^{I}_{AB}\vert BJ\rangle + \Gamma^{J}\vert BI\rangle\\
S_{A}\vert BI\rangle &=& \frac{1}{4}\Gamma^{J}_{AB}\vert IJ\rangle + \frac{1}{72}(\Gamma^{ILMN}_{AB} + 6\delta^{IL}\Gamma^{MN}_{AB})\vert LMN\rangle\\
S_{A}\vert LMN\rangle &=& \Gamma^{LM}_{AB}\vert BN\rangle + \Gamma^{MN}_{AB}\vert BL\rangle + \Gamma^{NL}_{AB}\vert BM\rangle
\end{eqnarray}
We can check that these definitions indeed reproduce the desired algebra. Let us check the statement explicitly for the graviton $\vert IJ\rangle$:
\begin{eqnarray*}
S_{A}S_{B}\vert IJ\rangle &=& \Gamma^{I}_{BC}S_{A}\vert CJ\rangle + \Gamma^{J}_{BC}S_{A}\vert CI\rangle\\
&=& \Gamma^{I}_{BC}[\frac{1}{4}\Gamma^{K}_{AC}\vert JK\rangle + \frac{1}{72}(\Gamma^{JLMN}_{AC} + 6\delta^{JL}\Gamma^{MN}_{AC})\vert LMN\rangle]\\ &+& \Gamma^{J}_{BC}[\frac{1}{4}\Gamma^{K}_{AC}\vert IK\rangle + \frac{1}{72}(\Gamma^{ILMN}_{AC} + 6\delta^{JL}\Gamma^{MN}_{AC})\vert LMN\rangle]
\end{eqnarray*}
Analogously,
\begin{eqnarray*}
S_{B}S_{A}\vert IJ\rangle &=& \Gamma^{I}_{AC}S_{B}\vert CJ\rangle + \Gamma^{J}_{AC}S_{B}\vert CI\rangle\\
&=& \Gamma^{I}_{AC}[\frac{1}{4}\Gamma^{K}_{BC}\vert JK\rangle + \frac{1}{72}(\Gamma^{JLMN}_{BC} + 6\delta^{JL}\Gamma^{MN}_{BC})\vert LMN\rangle]\\ &+& \Gamma^{J}_{AC}[\frac{1}{4}\Gamma^{K}_{BC}\vert IK\rangle + \frac{1}{72}(\Gamma^{ILMN}_{BC} + 6\delta^{JL}\Gamma^{MN}_{BC})\vert LMN\rangle]
\end{eqnarray*}
Thus, the anticommutator is
\begin{eqnarray*}
\{S_{A}, S_{B}\}\vert IJ\rangle &=& \frac{1}{4}[\Gamma^{I}_{BC}\Gamma^{K}_{AC} + \Gamma^{I}_{AC}\Gamma^{K}_{BC}]\vert JK\rangle + \frac{1}{4}[\Gamma^{J}_{BC}\Gamma^{K}_{AC} + \Gamma^{J}_{AC}\Gamma^{K}_{BC}]\vert IK\rangle\\
&+& \frac{1}{72}[(\Gamma^{I}_{BC}\Gamma^{JLMN}_{AC} + \Gamma^{J}_{BC}\Gamma^{ILMN}_{AC} + \Gamma^{I}_{AC}\Gamma^{JLMN}_{BC} + \Gamma^{J}_{AC}\Gamma^{ILMN}_{BC})\\
&+& 6(\delta^{JL}\Gamma^{I}_{BC}\Gamma^{MN}_{AC} + \delta^{IL}\Gamma^{J}_{BC}\Gamma^{MN}_{AC} + \delta^{JL}\Gamma^{I}_{AC}\Gamma^{MN}_{BC} + \delta^{IL}\Gamma^{J}_{AC}\Gamma^{MN}_{BC})]\vert LMN\rangle\\
&=& \frac{1}{4}(2\delta^{IK}\delta_{AB}\vert JK\rangle + 2\delta^{JK}\delta_{AB}\vert IK\rangle) + \frac{1}{72}[4!(\delta^{I[J}\Gamma^{LMN]}_{BA} + \delta^{J[I}\Gamma^{LMN]}_{BA} \\&+& \delta^{I[J}\Gamma^{LMN]}_{AB} + \delta^{J[I}\Gamma^{LMN]}_{AB}) + 6(\delta^{JL}\Gamma^{I}_{BC}\Gamma^{MN}_{AC} + \delta^{IL}\Gamma^{J}_{BC}\Gamma^{MN}_{AC}\\ &+& \delta^{JL}\Gamma^{I}_{AC}\Gamma^{MN}_{BC} + \delta^{IL}\Gamma^{J}_{AC}\Gamma^{MN}_{BC})]\vert LMN\rangle
\end{eqnarray*}
Now, let us consider the symmetry properties of the $SO(9)$ $\Gamma$-matrices. The 1-form and 4-form are symmetric in their spinor indices, and the 2-form and 3-form are antisymmetric in their spinor indices. Therefore,
\begin{eqnarray}
\{S_{A}, S_{B}\}\vert IJ\rangle &=& \delta_{AB}\vert IJ\rangle + \frac{1}{12}(\delta^{JL}\Gamma^{I}_{BC}\Gamma^{MN}_{AC} + \delta^{IL}\Gamma^{J}_{BC}\Gamma^{MN}_{AC}\nonumber\\
&+& \delta^{JL}\Gamma^{I}_{AC}\Gamma^{MN}_{BC} + \delta^{IL}\Gamma^{J}_{AC}\Gamma^{MN}_{BC})\vert LMN\rangle \nonumber\\
&=& \delta_{AB}\vert IJ\rangle + \frac{1}{12}[\delta^{JL}(\Gamma^{IMN}_{BA} + \delta^{I[M}\Gamma^{N]} + \Gamma^{IMN}_{AB} + \delta^{I[M}\Gamma^{N]}_{AB}) \nonumber\\
&+& \delta^{IL}(\Gamma^{JMN}_{BA} + \delta^{J[M}\Gamma^{N]} + \Gamma^{JMN}_{AB}  + \delta^{J[M}\Gamma^{N}_{AB})]\vert LMN\rangle\nonumber\\
&=& \delta_{AB}\vert IJ\rangle + \frac{1}{12}[\delta^{JL}\delta^{IM}\Gamma^{N} - \delta^{JL}\delta^{IN}\Gamma^{M} + \delta^{JL}\delta^{IM}\Gamma^{N} - \delta^{IN}\delta^{JL}\Gamma^{M}\\ &+& \delta^{IL}\delta^{JM}\Gamma^{N} - \delta^{IL}\delta^{JN}\Gamma^{M} + \delta^{IL}\delta^{JM}\Gamma^{N} - \delta^{IL}\delta^{JN}\Gamma^{M}]\vert LMN\rangle \nonumber\\
&=& \delta_{AB}\vert IJ\rangle + \frac{1}{12}[2\Gamma^{N}\vert JIN\rangle - 2\Gamma^{M}\vert JMI\rangle + 2\Gamma^{N}\vert IJN\rangle - 2\Gamma^{M}\vert IMJ\rangle]\nonumber\\
&=& \delta_{AB}\vert IJ\rangle
\end{eqnarray}
as expected. One can similarly show that this algebra is satisfied for the action of $S_{A}$ on the other two fields. Therefore, we have shown that the $D=11$ superparticle spectrum describes the physical degrees of freedom of $D=11$ supergravity.

\section{D=11 pure spinor superparticle}
As for the $D=10$ case \cite{Berkovits:2002zk}, we will obtain the $D=11$ pure spinor superparticle from the gauge-fixed Brink-Schwarz-like superparticle \eqref{eq72} by introducing a new set of variables $(\Theta^{\alpha}, P_{\alpha})$ and a new symmetry coming from the following first-class constraints:
\begin{equation}
\hat{D}_{\alpha} = D_{\alpha} + \frac{1}{\sqrt{\sqrt{2}P^{+}}}(\Gamma^{m}\Gamma^{+}S)_{\alpha}P_{m}
\end{equation}
where $\{S_{A}, S_{B}\} = \delta_{AB}$ and $D_{\alpha} = P_{\alpha} + \Gamma^{m}_{\alpha\beta}\Theta^{\beta}P_{m}$. Using the relation $\{\Theta^{\alpha}, P_{\beta}\} = i\delta^{\alpha}_{\beta}$, one can show that $\{D_{\alpha}, D_{\beta}\} = 2i(\Gamma^{m})_{\alpha\beta}P_{m}$. Let us check that these ones are indeed first-class constraints:
\begin{eqnarray}
\{\hat{D}_{\alpha}, \hat{D}_{\beta}\} &=& \{D_{\alpha}, D_{\beta}\} + \frac{1}{\sqrt{2}P^{+}}(\Gamma^{m})_{\alpha\lambda}(\Gamma^{+})^{\lambda A}(\Gamma^{n})_{\beta\delta}(\Gamma^{+})^{\delta A}P_{m}P_{n}\nonumber \\
&=& 2i(\Gamma^{m})_{\alpha\beta}P_{m} - \frac{\sqrt{2}i}{\sqrt{2}P^{+}}\Gamma^{m}_{\alpha\lambda}\Gamma^{n}_{\beta\delta}\Gamma^{+\,\lambda\delta}P_{m}P_{n}\nonumber \\
&=& 2i(\Gamma^{m})_{\alpha\beta}P_{m} - \frac{i}{P^{+}}(\Gamma^{m}\Gamma^{+}\Gamma^{n})_{\alpha\beta}P_{m}P_{n}
\end{eqnarray}
Since $\Gamma^{m}\Gamma^{+} = -\Gamma^{+}\Gamma^{m} + 2\eta^{m+}$, we obtain
\begin{eqnarray}
\{\hat{D}_{\alpha}, \hat{D}_{\beta}\} &=& 2i(\Gamma^{m})_{\alpha\beta}P_{m} - \frac{i}{P^{+}}(\Gamma^{m}\Gamma^{+}\Gamma^{n})_{\alpha\beta}P_{m}P_{n}\nonumber\\
&=& 2i(\Gamma^{m})_{\alpha\beta}P_{m} - \frac{i}{P^{+}}(2\eta^{+m}\Gamma^{n}_{\alpha\beta})P_{m}P_{n} + \frac{i}{P^{+}}\Gamma^{+}_{\alpha\beta}P^{2}\nonumber\\
&=& 2i(\Gamma^{m})_{\alpha\beta}P_{m} - 2i(\Gamma^{m})_{\alpha\beta}P_{m} + \frac{i}{P^{+}}\Gamma^{+}_{\alpha\beta}P^{2}\nonumber\\
&=& \frac{i}{P^{+}}\Gamma^{+}_{\alpha\beta}P^{2} \label{eq35}
\end{eqnarray}
Thus, the modified Brink-Schwarz-like action will be:
\begin{equation}
S = \int d\tau (\dot{X}^{m}P_{m} - \frac{i}{2}\dot{S}^{A}S_{A} + e P^{m}P_{m} + \dot{\Theta}^{\alpha}P_{\alpha} + f^{\alpha}\hat{D}_{\alpha})
\end{equation}
where we have added the usual kinetic term for the variables $(\Theta^{\alpha}, P_{\alpha})$ and the last term takes into account the new constraint through the fermionic Lagrange multiplier $f^{\alpha}$. The standard BRST method gives us the following gauge-fixed action: 
\begin{equation}\label{eq36}
S = \int d\tau(\dot{X}^{m}P_{m} - \frac{i}{2}\dot{S}^{A}S_{A} - \frac{1}{2}P^{m}P_{m} + \dot{\Theta}^{\alpha}P_{\alpha} + \dot{c}b + \dot{\hat{\Lambda}}^{\alpha}\hat{W}_{\alpha})
\end{equation}
and the BRST operator
\begin{equation}
\hat{Q} = \hat{\Lambda}^{\alpha}\hat{D}_{\alpha} + cP^{m}P_{m} - \frac{i}{2P^{+}}(\hat{\Lambda}\Gamma^{+}\hat{\Lambda})b,
\end{equation}
once we choose the gauge $e=-\frac{1}{2}$ and $f^{\alpha}=0$. The ghosts $c$, $\hat{\Lambda}^{\alpha}$ come from gauge-fixing the reparametrization symmetry and the new fermionic symmetry, respectively.

\vspace{2mm}
Now we will show that the cohomology of the BRST operator $\hat{Q}$ is equivalent to the cohomology of a BRST operator $Q = \Lambda^{\alpha}D_{\alpha}$, where $\Lambda^{\alpha}$ is a pure spinor. We will show this claim in two steps. First, we show that the $\hat{Q}$-cohomology is equivalent to $Q\ensuremath{'}$-cohomology, where $Q\ensuremath{'} = \Lambda\ensuremath{'}^{\alpha}\hat{D}_{\alpha}$ and $\Lambda\ensuremath{'}\Gamma^{+}\Lambda\ensuremath{'} = 0$. Finally, we will prove that the $Q\ensuremath{'}$-cohomology is equivalent to the $Q$-cohomology.

\vspace{2mm}
Let us start by defining the operator $Q_{0} = \Lambda_{0}^{\alpha}\hat{D}_{\alpha}$. Notice that when $\Lambda_{0}^{\alpha}$ is equal to $\hat{\Lambda}^{\alpha}$ or $\Lambda\ensuremath{'}^{\alpha}$, $Q_{0}$ becomes the first term of $\hat{Q}$ or $Q\ensuremath{'}$, respectively. Now, let $V$ be a state such that $Q_{0}V = (\Lambda_{0}\Gamma^{+}\Lambda_{0})W$, for some W. Because of the property that $\Lambda\ensuremath{'}^{\alpha}$ satisfies, V is annihilated by $Q\ensuremath{'}$. Also, using \eqref{eq35}, we find that $(Q_{0})^{2} = \frac{i}{2P^{+}}P^{m}P_{m}(\Lambda_{0}\Gamma^{+}\Lambda_{0})$. So, we conclude that $Q_{0}W = \frac{i}{2P^{+}}P^{m}P_{m}V$. We can then show that the state $\hat{V} = V - 2iP^{+}cW$ is annihilated by $\hat{Q}$:
\begin{eqnarray}
\hat{Q}\hat{V} &=& \hat{Q}(V - 2iP^{+}cW)\nonumber\\
 &=& \hat{Q}V - 2iP^{+}(\hat{Q}c)W + 2iP^{+}c(\hat{Q}W) \nonumber\\
 &=& (\hat{\Lambda}\Gamma^{+}\hat{\Lambda})W + c P^{m}P_{m}V - 2iP^{+}(-\frac{i}{2P^{+}})(\hat{\Lambda}\Gamma^{+}\hat{\Lambda})W + 2iP^{+}c(\frac{i}{2P^{+}})P^{m}P_{m}V \nonumber\\
 &=& (\hat{\Lambda}\Gamma^{+}\hat{\Lambda})W + c P^{m}P_{m}V - (\hat{\Lambda}\Gamma^{+}\hat{\Lambda})W - c P^{m}P_{m}V\nonumber\\
 &=& 0
\end{eqnarray}
where we have assumed that $b$ annihilates physical states. Now, let us show that if a state $V$ is BRST-trivial (in the $Q\ensuremath{'}$-cohomology), we can find a state $\hat{V} = V - 2iP^{+}cW$ which is also BRST-trivial (in the $\hat{Q}$-cohomology). Let V be a state which satisfies $V = Q_{0}\Omega + (\Lambda_{0}\Gamma^{+}\Lambda_{0})Y$, for some $Y$. It is clear that if $\Lambda_{0}^{\alpha} = \Lambda\ensuremath{'}^{\alpha}$, we have that $V$ is $Q\ensuremath{'}$-exact and if $\Lambda_{0}^{\alpha} = \hat{\Lambda}^{\alpha}$, we have that the first term of $\hat{Q}$ is equal to $V - (\hat{\Lambda}\Gamma^{+}\hat{\Lambda})Y$. So we see that
\begin{eqnarray}
\hat{Q}(\Omega + 2iP^{+}cY) &=& \hat{Q}\Omega + 2iP^{+}(\hat{Q}c)Y - 2iP^{+}c(\hat{Q}Y)\nonumber\\
 &=& V - (\hat{\Lambda}\Gamma^{+}\hat{\Lambda})Y + cP^{m}P_{m} + 2iP^{+}(-\frac{i}{2P^{+}})(\hat{\Lambda}\Gamma^{+}\hat{\Lambda})Y \nonumber\\
 && -
2iP^{+}c(W - \frac{i}{2P^{+}}P^{m}P_{m}\Omega)
\end{eqnarray}
where we used the fact that $b$ annihilates $\Omega$ as well as the result $\hat{Q}_{0}Y =
W - \frac{i}{2P^{+}}P^{m}P_{m}$, which follows from the definition of $V$. Hence, we obtain
\begin{eqnarray}
\hat{Q}(\Omega + 2iP^{+}cY) &=& V - (\hat{\Lambda}\Gamma^{+}\hat{\Lambda})Y + cP^{m}P_{m}\Omega + (\hat{\Lambda}\Gamma^{+}\hat{\Lambda})Y - 2iP^{+}cW - cP^{m}P_{m}\Omega\nonumber\\
 &=& V - 2iP^{+}cW\nonumber\\
 &=& \hat{V}
\end{eqnarray}
Therefore, we have proven that for each state $V$ in the $Q\ensuremath{'}$-cohomology, we can find a state $\hat{V}$ in the $\hat{Q}$-cohomology. If we reverse the arguments given above we can show that any state in the $\hat{Q}$-cohomology corresponds to a state in the $Q\ensuremath{'}$-cohomology.

\vspace{2mm}
The last step is to show that the $Q\ensuremath{'}$-cohomology is equivalent to the $Q$-cohomology. We will do this by using two different approaches.

\subsection{Group decomposition $SO(9) \rightarrow SU(2)\times SU(4)$}
The $SO(10,1)$ spinors $\Lambda^{\alpha}$ and $D_{\alpha}$ can be expressed in terms of their $SO(8)$ components in the following way:
\begin{equation}\label{eq500}
\begin{aligned}
\Lambda\ensuremath{'}^{\alpha} &= \begin{pmatrix}
\lambda\ensuremath{'}^{a}\\
\lambda\ensuremath{'}^{\dot{a}}\\
\tilde{\lambda}\ensuremath{'}^{a}\\
\tilde{\lambda}\ensuremath{'}^{\dot{a}}
\end{pmatrix}
\end{aligned}, \hspace{4mm}
\begin{aligned}
D_{\alpha} &=& \begin{pmatrix}
\tilde{d}^{a}\\
\tilde{d}^{\dot{a}}\\
-d^{a}\\
-d^{\dot{a}}
\end{pmatrix}
\end{aligned},
\end{equation}
where $a, \dot{a} = 1,\ldots, 8$. The constraint $\Lambda\ensuremath{'}\Gamma^{+}\Lambda\ensuremath{'} = 0$ can be written in terms of these $SO(8)$ components as follows 
\begin{equation}
\lambda\ensuremath{'}^{\dot{a}}\lambda\ensuremath{'}^{\dot{a}} + \tilde{\lambda}\ensuremath{'}^{a}\tilde{\lambda}\ensuremath{'}^{a} = 0
\end{equation}
The particular representation for $SO(10,1)$ $\Gamma$-matrices used in this section is studied in detail in Appendix \ref{appA}. Now, we find it useful to break $SO(9)$ into $SU(2)\times SU(4)$. The branching rule for the spinor representation is  $16 \rightarrow (2,4)+(2,\bar{4})$. Explicit expressions for the $SU(2)\times SU(4)$ components corresponding to $S^{a}, \bar{S}^{\dot{a}}, d^{\dot{a}}, \tilde{d}^{a}, \lambda\ensuremath{'}^{a}, \tilde{\lambda}\ensuremath{'}^{\dot{a}}$ are given below:
\begin{equation}\label{eq400}
\begin{aligned}
S_{\hat{A}} &= \frac{1}{\sqrt{2}}(S^{2a} + iS^{2a-1})\\
S_{\bar{\hat{A}}} &= \frac{1}{\sqrt{2}}(S^{2a} - iS^{2a-1})\\
\tilde{S}_{\hat{A}} &= \frac{1}{\sqrt{2}}(\bar{S}^{2\dot{a}} + i\bar{S}^{2\dot{a}-1})\\
\tilde{S}_{\bar{\hat{A}}} &= \frac{1}{\sqrt{2}}(\bar{S}^{2\dot{a}} - i\bar{S}^{2\dot{a}-1})
\end{aligned}\hspace{4mm}
\begin{aligned}
d_{\hat{A}} &= \frac{1}{\sqrt{2}}(d^{2\dot{a}} + id^{2\dot{a}-1})\\
d_{\bar{\hat{A}}} &= \frac{1}{\sqrt{2}}(d^{2\dot{a}} - id^{2\dot{a}-1})\\
\tilde{d}_{\hat{A}} &= \frac{1}{\sqrt{2}}(\tilde{d}^{2a} + i\tilde{d}^{2a-1})\\
\tilde{d}_{\bar{\hat{A}}} &= \frac{1}{\sqrt{2}}(\tilde{d}^{2a} - i\tilde{d}^{2a-1})
\end{aligned}\hspace{4mm}
\begin{aligned}
\lambda\ensuremath{'}_{\hat{A}} &= \frac{1}{\sqrt{2}}(\lambda\ensuremath{'}^{2a} + i\lambda\ensuremath{'}^{2a-1})\\
\lambda\ensuremath{'}_{\bar{\hat{A}}} &= \frac{1}{\sqrt{2}}(\lambda\ensuremath{'}^{2a} - i\lambda\ensuremath{'}^{2a-1})\\
\tilde{\lambda}\ensuremath{'}_{\hat{A}} &= \frac{1}{\sqrt{2}}(\tilde{\lambda}\ensuremath{'}^{2\dot{a}} + i\tilde{\lambda}\ensuremath{'}^{2\dot{a}-1})\\
\tilde{\lambda}\ensuremath{'}_{\bar{\hat{A}}} &= \frac{1}{\sqrt{2}}(\tilde{\lambda}\ensuremath{'}^{2\dot{a}} - i\tilde{\lambda}\ensuremath{'}^{2\dot{a}-1})
\end{aligned}
\end{equation}
where the $SO(9)$ spinor $S_{A}$ has been expressed in terms of its $SO(8)$ components:
\begin{equation}
S_{A} = \begin{pmatrix}
S^{a}\\
\bar{S}^{\dot{a}}
\end{pmatrix}
\end{equation}
and $\hat{A}, \bar{\hat{A}} = 1, \ldots , 4$. It should be clear in \eqref{eq400} that fields in the same representation of $SU(4)$ ($4$ or $\bar{4}$) form $SU(2)$ doublets. So, for instance, $\begin{pmatrix}
d_{\hat{A}}\\
\tilde{d}_{\hat{A}}
\end{pmatrix}$ transforms under $(2,4)$, $\begin{pmatrix}
\lambda\ensuremath{'}_{\bar{\hat{A}}}\\
\tilde{\lambda}\ensuremath{'}_{\bar{\hat{A}}}
\end{pmatrix}$ transforms under $(2,\bar{4})$, etc. Notice that the representations $4$ and $\bar{4}$ are defined by the null spinor $(\Gamma^{+}\Lambda\ensuremath{'})^{A}$ by using the fact that one can always choose an $SU(4)$ subgroup under which this spinor is invariant. Therefore we define the antifundamental representation ($\bar{4}$) in such a way that $(\Gamma^{J})_{(\Upsilon\bar{\hat{A}})A}(\Gamma^{+}\Lambda\ensuremath{'})^{A} = 0$, where $J=1,\ldots ,9$, $\Upsilon$ is an $SU(2)$ vector index and $A$ is an $SO(9)$ spinor index. After making the following shifts:
\begin{eqnarray}
S_{\hat{A}} &\rightarrow & S_{\hat{A}} - (\frac{\sqrt{\sqrt{2}}}{2\sqrt{P^{+}}})\tilde{d}_{\hat{A}}\\
\tilde{S}_{\hat{A}} &\rightarrow & \tilde{S}_{\hat{A}} + (\frac{\sqrt{\sqrt{2}}}{2\sqrt{P^{+}}})d_{\hat{A}}
\end{eqnarray}
the operator $Q\ensuremath{'}$ will change by the similarity transformation:
\begin{equation}
Q\ensuremath{'} \rightarrow e^{-[K(S_{\bar{\hat{A}}}\tilde{d}_{\hat{A}} - \tilde{S}_{\bar{\hat{A}}}d_{\hat{A}})]}Q\ensuremath{'}e^{[K(S_{\bar{\hat{A}}}\tilde{d}_{\hat{A}} - \tilde{S}_{\bar{\hat{A}}}d_{\hat{A}})]}
\end{equation}
where $K = -\frac{\sqrt{\sqrt{2}}}{2\sqrt{P{{+}}}}$. This result can be expanded by using the BCH formula:
\begin{equation}\label{eq402}
e^{-Z}Xe^{Z} = X + [X, Z] + \frac{1}{2}[[X, Z], Z] + \ldots
\end{equation}
where $X = Q\ensuremath{'} = \Lambda\ensuremath{'}^{\alpha}\hat{D}_{\alpha}$ and $Z = K(S_{\bar{\hat{A}}}\tilde{d}_{\hat{A}} - \tilde{S}_{\bar{\hat{A}}}d_{\hat{A}})$. The first term is just $Q\ensuremath{'}$,  which can be cast as
\begin{eqnarray}
Q\ensuremath{'} &=& \Lambda\ensuremath{'}^{\alpha}D_{\alpha} + \frac{1}{\sqrt{\sqrt{2}P^{+}}}(\Lambda\ensuremath{'}\Gamma^{m}\Gamma^{+}S) P_{m} \nonumber\\
&=& \lambda\ensuremath{'}_{\dot{a}}\tilde{d}_{\dot{a}} + \lambda\ensuremath{'}_{\bar{\hat{A}}}\tilde{d}_{\hat{A}} + \lambda\ensuremath{'}_{\hat{A}}\tilde{d}_{\bar{\hat{A}}} - \tilde{\lambda}\ensuremath{'}_{a}d_{a} - \tilde{\lambda}\ensuremath{'}_{\bar{\hat{A}}}d_{\hat{A}} - \tilde{\lambda}\ensuremath{'}_{\hat{A}}d_{\bar{\hat{A}}} + \sqrt{2\sqrt{2}P^{+}}\lambda\ensuremath{'}_{\bar{\hat{A}}}S_{\hat{A}} + \sqrt{2\sqrt{2}P^{+}}\lambda\ensuremath{'}_{\hat{A}}S_{\bar{\hat{A}}}\nonumber\\
&& + \sqrt{2\sqrt{2}P^{+}}\tilde{\lambda}\ensuremath{'}_{\bar{\hat{A}}}\tilde{S}_{\hat{A}} + \sqrt{2\sqrt{2}P^{+}}\tilde{\lambda}\ensuremath{'}_{\hat{A}}\tilde{S}_{\bar{\hat{A}}}+ \sqrt{\frac{\sqrt{2}}{P^{+}}}[\lambda\ensuremath{'}^{\dot{a}}(\sigma^{\hat{i}})_{\dot{a}\hat{A}}S_{\bar{\hat{A}}}P_{\hat{i}} - \tilde{\lambda}\ensuremath{'}^{a}(\sigma^{\hat{i}})_{\hat{A}a}\tilde{S}_{\bar{\hat{A}}}P_{\hat{i}}]\nonumber\\
&& + \sqrt{\frac{\sqrt{2}}{P^{+}}}[\tilde{\lambda}\ensuremath{'}_{\hat{A}}S_{\bar{\hat{A}}} + \lambda\ensuremath{'}_{\hat{A}}\tilde{S}_{\bar{\hat{A}}}]P^{11}\nonumber\\\label{eq107}
\end{eqnarray}
To find the second term in \eqref{eq402}, it is necessary to compute the $SU(4)$ (anti)commutation relations, which can be obtained from the $SO(8)$ relations:
\begin{equation}
\begin{aligned}
\{\tilde{d}^{a}, \tilde{d}^{b}\} &= -2\sqrt{2}\delta^{ab}P^{+} \hspace{4mm}, \\ 
\{d^{\dot{a}}, d^{\dot{b}}\} &= -2\sqrt{2}\delta^{\dot{a}\dot{b}}P^{+} \hspace{4mm} ,\\
\{d^{a}, d^{b}\} &= -2\sqrt{2}\delta^{ab}P^{-} \hspace{4mm}, \\ 
\{\tilde{d}^{\dot{a}}, \tilde{d}^{\dot{b}}\} &= -2\sqrt{2}\delta^{\dot{a}\dot{b}}P^{-} \hspace{4mm} ,
\end{aligned}\hspace{4mm}
\begin{aligned}
\{d^{a}, \tilde{d}^{b}\} &= 2\delta^{ab}P^{11}\\ 
\{\tilde{d}^{\dot{a}}, d^{\dot{b}}\} &= 2\delta^{\dot{a}\dot{b}}P^{11} \\
\{d^{a}, d^{\dot{b}}\} &= 2(\sigma^{\hat{i}})^{a\dot{b}}P^{\hat{i}}  \\ 
\{\tilde{d}^{\dot{a}}, d^{b}\} &= -2(\sigma^{\hat{i}})^{\dot{a}b}P^{\hat{i}} 
\end{aligned}
\end{equation}
Using these, together with \eqref{eq400}, leads us to the following $SU(4)$ relations:
\begin{equation}
\begin{aligned}
\{S_{\hat{A}}, S_{\bar{\hat{A}}}\} &= \eta_{\hat{A}\bar{\hat{A}}} \hspace{16.3mm}, \\ 
\{\tilde{S}_{\hat{A}}, \tilde{S}_{\bar{\hat{A}}}\} &= \eta_{\hat{A}\bar{\hat{A}}} \hspace{16.3mm} ,\\
\{d_{\hat{A}}, d_{\bar{\hat{A}}}\} &= -2\sqrt{2}\eta_{\hat{A}\bar{\hat{A}}}P^{+}\hspace{1mm}, \\
\{\tilde{d}_{\hat{A}}, \tilde{d}_{\bar{\hat{A}}}\} &= -2\sqrt{2}\eta_{\hat{A}\bar{\hat{A}}}P^{+}\hspace{1mm} ,
\end{aligned}
\begin{aligned}
\{ \tilde{d}_{\dot{a}}, d_{\hat{A}}\} &= 2\delta_{\dot{a}A}P^{11} \hspace{4mm} ,\\
\{ \tilde{d}_{\dot{a}}, d_{\bar{\hat{A}}}\} &= 2\delta_{\dot{a}\bar{A}}P^{11}\hspace{4mm} ,\\
\{ d_{a}, \tilde{d}_{\hat{A}}\} &= 2\delta_{a A}P^{11} \hspace{4mm} ,\\
\{ d_{a}, \tilde{d}_{\bar{\hat{A}}}\} &= 2\delta_{a \bar{A}}P^{11}\hspace{4mm} ,
\end{aligned}
\begin{aligned}
\{d_{a}, d_{\hat{A}}\} &=2(\sigma^{\hat{i}})_{a \hat{A}}P^{\hat{i}}\\
\{d_{a}, d_{\bar{\hat{A}}}\} &=2(\sigma^{\hat{i}})_{a \bar{\hat{A}}}P^{\hat{i}}\\
\{\tilde{d}_{\dot{a}}, \tilde{d}_{\hat{A}}\} &=-2(\sigma^{\hat{i}})_{\dot{a} \hat{A}}P^{\hat{i}}\\
\{\tilde{d}_{\dot{a}}, \tilde{d}_{\bar{\hat{A}}}\} &=-2(\sigma^{\hat{i}})_{\dot{a} \bar{\hat{A}}}P^{\hat{i}}
\end{aligned}
\end{equation}

Hence, we get
\begin{eqnarray}
K\left[Q\ensuremath{'}, S_{\bar{\hat{A}}}\tilde{d}_{\hat{A}} - \tilde{S}_{\bar{\hat{A}}}d_{\hat{A}}\right] &=& 
 -\sqrt{2\sqrt{2}P^{+}}\lambda\ensuremath{'}_{\hat{A}}S_{\bar{\hat{A}}} -\sqrt{2\sqrt{2}P^{+}}\tilde{\lambda}\ensuremath{'}_{\hat{A}}\tilde{S}_{\bar{\hat{A}}} - \lambda\ensuremath{'}_{\bar{\hat{A}}}\tilde{d}_{\hat{A}} + \tilde{\lambda}\ensuremath{'}_{\bar{\hat{A}}} d_{\hat{A}} -\sqrt{\frac{\sqrt{2}}{P^{+}}}\tilde{\lambda}\ensuremath{'}_{\hat{A}}S_{\bar{\hat{A}}}P^{11}\nonumber\\
 && -\sqrt{\frac{\sqrt{2}}{P^{+}}}\lambda\ensuremath{'}_{\hat{A}}\tilde{S}_{\bar{\hat{A}}}P^{11} -\sqrt{\frac{\sqrt{2}}{P^{+}}}\lambda\ensuremath{'}^{\dot{a}}(\sigma^{\hat{i}})_{\dot{a}\hat{A}}S_{\bar{\hat{A}}}P_{\hat{i}} + \sqrt{\frac{\sqrt{2}}{P^{+}}}\tilde{\lambda}\ensuremath{'}^{a}(\sigma^{\hat{i}})_{\hat{A} a}\tilde{S}_{\bar{A}}P_{\hat{i}}\label{eq106}
\end{eqnarray}
From this expression it is easy to see that:
\begin{equation}
[[Q\ensuremath{'}, Z], Z] = 0
\end{equation}
and so the third term and all of the other ones in \eqref{eq402} (which were represented by $\ldots$) vanish.

\vspace{2mm}
Therefore, we have arrived at the following result:
\begin{equation}
Q\ensuremath{'} \rightarrow \lambda\ensuremath{'}_{\dot{a}}\tilde{d}_{\dot{a}} + \lambda\ensuremath{'}_{\hat{A}}\tilde{d}_{\bar{\hat{A}}} - \tilde{\lambda}\ensuremath{'}_{a}d_{a} - \tilde{\lambda}\ensuremath{'}_{\hat{A}}d_{\bar{\hat{A}}} + \sqrt{2\sqrt{2}P^{+}}\lambda\ensuremath{'}_{\bar{\hat{A}}}S_{\hat{A}} + \sqrt{2\sqrt{2}P^{+}}\tilde{\lambda}\ensuremath{'}_{\bar{\hat{A}}}\tilde{S}_{\hat{A}}
\end{equation}
where $\lambda\ensuremath{'}^{\dot{a}}$ and $\tilde{\lambda}\ensuremath{'}^{a}$ satisfy the relation $\lambda\ensuremath^{\dot{a}}\lambda\ensuremath^{\dot{a}} + \tilde{\lambda}\ensuremath{'}^{a}\tilde{\lambda}\ensuremath{'}^{a} = 0$. If we define a spinor $\Lambda^{\alpha} = [\lambda_{\hat{A}}, \lambda_{\bar{\hat{A}}}, \lambda_{\dot{a}}, \tilde{\lambda}_{a}, \tilde{\lambda}_{\hat{A}}, \tilde{\lambda}_{\bar{\hat{A}}}] = [\lambda\ensuremath{'}_{\hat{A}}, 0, \lambda\ensuremath{'}_{\dot{a}}, \tilde{\lambda}\ensuremath{'}_{a}, \tilde{\lambda}\ensuremath{'}_{\hat{A}}, 0]$, the previous expression can be written as
\begin{equation}
Q\ensuremath{'} \rightarrow \Lambda^{\alpha}D_{\alpha} + \sqrt{2\sqrt{2}P^{+}}\lambda\ensuremath{'}_{\bar{\hat{A}}}S_{\hat{A}} + \sqrt{2\sqrt{2}P^{+}}\tilde{\lambda}\ensuremath{'}_{\bar{\hat{A}}}\tilde{S}_{\hat{A}}
\end{equation} 
Furthermore, after using the quartet argument \cite{Kugo:1979gm}, it is clear that the $Q\ensuremath{'}$-cohomology is equivalent to the $Q$-cohomology\footnote{That is, the states in the Hilbert space will be independent of $\lambda\ensuremath{'}_{\bar{\hat{A}}}$, $S_{\hat{A}}$, $\tilde{\lambda}\ensuremath{'}_{\bar{\hat{A}}}$, $\tilde{S}_{\hat{A}}$, and their respective conjugate momenta $w\ensuremath{'}_{\hat{A}}$, $S_{\bar{\hat{A}}}$, $\tilde{w}\ensuremath{'}_{\hat{A}}$, $\tilde{S}_{\bar{\hat{A}}}$.}:
\begin{equation}
Q\ensuremath{'} \rightarrow Q = \Lambda^{\alpha}D_{\alpha}
\end{equation}
where $\Lambda^{\alpha}$ is a pure spinor.

\subsection{Group decomposition $SO(9) \rightarrow U(1) \times SO(7)$}
We will express $SO(10,1)$ spinors in terms of their $SO(3,1) \times SO(7)$ components:
\begin{equation}
\chi^{\alpha} = \begin{pmatrix}
\chi^{\pm\pm 0}\\
\chi^{\pm\pm i}
\end{pmatrix}
\end{equation}
where $i = 1, \ldots, 7$. The notation $\pm$ and the representation of the $SO(10,1)$ gamma matrices used here are explained in detail in Appendix \ref{appB}. Using this notation, we can express the (anti)commutation relations studied above in the $SO(3,1) \times SO(7)$ language:
\begin{equation}\label{eq405}
\begin{aligned}
\{D^{--0}, D^{-+0}\} &= 2\sqrt{2}P^{+}\hspace{9.2mm},\\ 
\{D^{--i}, D^{-+j}\} &= -2\sqrt{2}P^{+}\delta^{ij}\hspace{2mm},\\
\{D^{++0}, D^{+-0}\} &= 2\sqrt{2}P^{-}\hspace{9.4mm},\\ 
\{D^{++i}, D^{+-j}\} &= -2\sqrt{2}P^{-}\delta^{ij}\hspace{2.1mm},
\end{aligned}
\begin{aligned}
\{D^{--0}, D^{+-0}\} &= 2\sqrt{2}P^{2+3i}\hspace{9.2mm},\\
\{D^{--i}, D^{+-j}\} &= -2\sqrt{2}P^{2+3i}\delta^{ij}\hspace{2mm},\\
\{D^{++0}, D^{-+0}\} &= 2\sqrt{2}P^{2-3i}\hspace{9.4mm},\\
\{D^{++i}, D^{-+j}\} &= -2\sqrt{2}P^{2-3i}\delta^{ij}\hspace{2.1mm},
\end{aligned}
\begin{aligned}
\{D^{--i}, D^{++0}\} &= -2P^{i}\\
\{D^{++i}, D^{--0}\} &= 2P^{i}\\
\{D^{-+i}, D^{+-0}\} &= 2P^{i}\\
\{D^{+-i}, D^{-+0}\} &= -2P^{i}\\
\end{aligned}
\end{equation}
and also
\begin{equation}
\begin{aligned}
\{S^{--0}, S^{-+0}\} &= -1 \\
\{S^{--i}, S^{-+j}\} &= \delta^{ij} 
\end{aligned}
\end{equation}
and any other anticommutator vanishes. Under a certain subgroup $U(1)\times SO(7) \subset SO(9)$, the null spinor $(\Gamma^{+}\Lambda\ensuremath{'})^{A}$ will be invariant up to rescaling. This subgroup is chosen in such a way that $(\Gamma^{2+3i})_{(-0)A}(\Gamma^{+}\Lambda\ensuremath{'})^{A}$ 
$ = (\Gamma^{j})_{(-0)A}(\Gamma^{+}\Lambda\ensuremath{'})^{A} = 0$
, where we have dropped out the minus sign associated to the first $U(1)$ charge, and $j=1,\ldots ,7$.
\vspace{2mm}
The BRST operator $Q\ensuremath{'}$ can be expressed in terms of $SO(3,1)\times SO(7)$ variables:
\begin{eqnarray}
Q\ensuremath{'} &=& \Lambda\ensuremath{'}^{\alpha}D_{\alpha} + \frac{1}{\sqrt{\sqrt{2}P^{+}}}[-(\Lambda\ensuremath{'}\Gamma^{-}\Gamma^{+}S)P^{+} +  (\Lambda\ensuremath{'}\Gamma^{2-3i}\Gamma^{+}S)P^{2+3i} +  (\Lambda\ensuremath{'}\Gamma^{2+3i}\Gamma^{+}S)P^{2-3i}\nonumber\\
&& + (\Lambda\ensuremath{'}\Gamma^{j}\Gamma^{+}S)P^{j}]\nonumber\\
&=& \Lambda\ensuremath{'}^{\alpha}D_{\alpha} - \frac{2\sqrt{P^{+}}}{\sqrt{\sqrt{2}}}(\Lambda\ensuremath{'}^{+-0}S^{-+0} - \Lambda\ensuremath{'}^{+-i}S^{-+i} + \Lambda\ensuremath{'}^{++0}S^{--0} - \Lambda\ensuremath{'}^{++i}S^{--i})\nonumber\\
&& +\sqrt{\frac{2\sqrt{2}}{P^{+}}}(\Lambda^{-+0}S^{-+0})P^{2+3i} - \sqrt{\frac{2\sqrt{2}}{P^{+}}}(\Lambda^{--j}S^{--j})P^{2-3i}-\sqrt{\frac{\sqrt{2}}{P^{+}}}(\Lambda^{--j}S^{-+0})P^{j}\nonumber\\
&& -\sqrt{\frac{\sqrt{2}}{P^{+}}}(\Lambda^{-+0}S^{--j})P^{j}\nonumber
\end{eqnarray}

After performing the following shifts:
\begin{eqnarray}
S^{--0} \rightarrow S^{--0} - \frac{\sqrt{\sqrt{2}}}{2\sqrt{P^{+}}}D^{--0}\\
S^{-+i} \rightarrow S^{-+i} - \frac{\sqrt{\sqrt{2}}}{2\sqrt{P^{+}}}D^{-+i}
\end{eqnarray}
the BRST operator will changed by
\begin{equation}
Q\ensuremath{'} \rightarrow e^{-Z}Q\ensuremath{'}e^{Z},
\end{equation}
where $Z = \frac{\sqrt{\sqrt{2}}}{2\sqrt{P^{+}}}(S^{-+0}D^{--0} - S^{--i}D^{-+i})$. The BCH formula \eqref{eq402} gives us the result
\begin{eqnarray}
Q\ensuremath{'} &\rightarrow & Q\ensuremath{'} + [Q\ensuremath{'}, Z] + \frac{1}{2}[[Q\ensuremath{'}, Z], Z] + \ldots\nonumber\\
&\rightarrow & -\Lambda\ensuremath{'}^{++0}D^{--0} + \Lambda\ensuremath{'}^{++i}D^{--i} + \Lambda\ensuremath{'}^{--0}D^{++0} - \Lambda\ensuremath{'}^{--i}D^{++i} -\Lambda\ensuremath{'}^{+-0}D^{-+0} + \Lambda\ensuremath{'}^{+-i}D^{-+i} \nonumber\\
&&+ \Lambda\ensuremath{'}^{-+0}D^{+-0} - \Lambda\ensuremath{'}^{-+i}D^{+-i} - \frac{2\sqrt{P^{+}}}{\sqrt{\sqrt{2}}}(\Lambda\ensuremath{'}^{+-0}S^{-+0} - \Lambda\ensuremath{'}^{++i}S^{--i} + \Lambda\ensuremath{'}^{++0}S^{--0} \nonumber\\
&&- \Lambda\ensuremath{'}^{+-i}S^{-+i}) +\sqrt{\frac{2\sqrt{2}}{P^{+}}}(\Lambda^{-+0}S^{-+0})P^{2+3i} - \sqrt{\frac{2\sqrt{2}}{P^{+}}}(\Lambda^{--j}S^{--j})P^{2-3i}\nonumber\\
&&-\sqrt{\frac{\sqrt{2}}{P^{+}}}(\Lambda^{--j}S^{-+0})P^{j}-\sqrt{\frac{\sqrt{2}}{P^{+}}}(\Lambda^{-+0}S^{--j})P^{j}+ \frac{\sqrt{\sqrt{2}}}{2\sqrt{P^{+}}}[2\sqrt{2}\Lambda\ensuremath{'}^{+-0}S^{-+0}P^{+}\nonumber\\
&& + 2\Lambda\ensuremath{'}^{--i}S^{-+0}P^{i} - 2\sqrt{2}\Lambda\ensuremath{'}^{-+0}S^{-+0}P^{2+3i}  - 2\sqrt{2}P^{+}\Lambda\ensuremath{'}^{++i}S^{--i} + 2\sqrt{2}P^{2-3i}\Lambda\ensuremath{'}^{--i}S^{--i} \nonumber\\
&&+ 2\Lambda\ensuremath{'}^{-+0}S^{--i}P^{i}] + \Lambda\ensuremath{'}^{++0}D^{--0} -\Lambda\ensuremath{'}^{+-i}D^{-+i}+\ldots
\end{eqnarray}
where the ellipsis represents $\frac{1}{2!}[[Q\ensuremath{'}, Z], Z] + \frac{1}{3!}[[[Q\ensuremath{'}, Z], Z], Z] + \ldots$. However, these terms vanish because $[[Q\ensuremath{'}, Z], Z] = 0$, as can be seen from the equation \eqref{eq405}. Thus, we are left with
\begin{eqnarray}
Q\ensuremath{'}&\rightarrow & \Lambda\ensuremath{'}^{++i}D^{--i} + \Lambda\ensuremath{'}^{--0}D^{++0} - \Lambda\ensuremath{'}^{--i}D^{++i} - \Lambda\ensuremath{'}^{+-0}D^{-+0} + \Lambda\ensuremath{'}^{-+0}D^{+-0} - \Lambda\ensuremath{'}^{-+i}D^{+-i}\nonumber\\
&& -\sqrt{2\sqrt{2}P^{+}}(\Lambda\ensuremath{'}^{++0}S^{--0} - \Lambda\ensuremath{'}^{+-i}S^{-+i})
\end{eqnarray}
If we define a spinor $\Lambda^{\alpha} = [\Lambda^{++0}, \Lambda^{++i}, \Lambda^{--0}, \Lambda^{--i}, \Lambda^{+-0}, \Lambda^{+-i}, \Lambda^{-+0}, \Lambda^{-+i}]$ $=$ $[0$, $ \Lambda\ensuremath{'}^{++i}$, $\Lambda\ensuremath{'}^{--0}$, $\Lambda\ensuremath{'}^{--i}$,$\Lambda\ensuremath{'}^{+-0}$, $0$, $\Lambda\ensuremath{'}^{-+0}$, $\Lambda\ensuremath{'}^{-+i}]$ where $\Lambda\ensuremath{'}\Gamma^{+}\Lambda\ensuremath{'} = 0$, the resulting BRST operator can be written as
\begin{equation}\label{eq410}
Q\ensuremath{'} \rightarrow \Lambda^{\alpha}D_{\alpha} - \sqrt{2\sqrt{2}P^{+}}(\Lambda\ensuremath{'}^{++0}S^{-+0} - \Lambda\ensuremath{'}^{+-i}S^{--i})
\end{equation}
From this last expression, we can conclude that the space of physical states will not depend on the canonical variables $S^{-+0}$, $S^{--i}$, $\Lambda\ensuremath{'}^{++0}$, $\Lambda\ensuremath{'}^{+-i}$, or their respective conjugate momenta $S^{--0}$, $S^{-+i}$, $W\ensuremath{'}^{--0}$, $W\ensuremath{'}^{-+i}$. Therefore the BRST operator takes the simple form
\begin{equation}
Q\ensuremath{'} \rightarrow Q = \Lambda^{\alpha}D_{\alpha}
\end{equation}
where $\Lambda^{\alpha}$ is a $D=11$ pure spinor. Therefore, we have proved that the modified Brink-Schwarz like superparticle action \eqref{eq36} is equivalent to the theory described by the manifestly Lorentz covariant action
\begin{equation}
S = \int d\tau (\dot{X}^{m}P_{m} - \frac{1}{2}P^{m}P_{m} + \dot{\Theta}^{\alpha}P_{\alpha} + \dot{\Lambda}^{\alpha}W_{\alpha})
\end{equation}
and the BRST operator $Q = \Lambda^{\alpha}D_{\alpha}$, where $\Lambda\Gamma^{m}\Lambda = 0$. This theory is the $D=11$ pure spinor superparticle.

\section{Light-cone analysis of the pure spinor cohomology}
In this section it will be shown that the pure spinor physical condition implies light-cone equations of motion for $D=11$ linearized supergravity in $D=9$ superspace, which coincide with those found in \cite{Green:1999by}. To see this, let us write $Q$ in $SO(9)$ notation (see Appendix \ref{appA}):
\begin{equation}
Q = \Lambda^{A}D_{A} + \bar{\Lambda}^{A}\bar{D}_{A}
\end{equation}
and define the operator
\begin{equation}
R = \frac{P^{I}\bar{N}_{I}}{\sqrt{2}P^{+}}
\end{equation}
where $I=1,\ldots,8,11$ and $\bar{N}^{I} = \Lambda^{A}\Gamma^{I}_{AB}\bar{W}^{B}$. The corresponding similarity transformation generated by this operator is
\begin{eqnarray}
\tilde{Q} &=& e^{-R}Qe^{R} \nonumber\\
&=& Q + [Q,R] + \frac{1}{2}[[Q,R],R] + \ldots \nonumber\\
&=& \Lambda^{A}D_{A} + \bar{\Lambda}^{A}\bar{D}_{A} + \frac{i}{\sqrt{2}P^{+}}P_{I}(\Lambda^{A}\Gamma^{I}_{AB}\bar{D}^{B})\nonumber\\
&=& \Lambda^{A}[D_{A} + \frac{i}{\sqrt{2}P^{+}}P_{I}(\Gamma^{I}\bar{D})_{A}] + \bar{\Lambda}^{A}\bar{D}_{A}\nonumber\\
&=& \Lambda^{A}G_{A} + \bar{\Lambda}^{A}\bar{D}_{A}
\end{eqnarray}
where $G_{A}$ is defined by the relation
\begin{eqnarray}
G_{A} &=& D_{A} + \frac{i}{\sqrt{2}P^{+}}P_{I}(\Gamma^{I}\bar{D})_{A}\nonumber\\
&=& D_{A} + \frac{1}{\sqrt{2}P^{+}}P_{\hat{i}}(\gamma^{9}\gamma^{\hat{i}}\bar{D})_{A} - \frac{1}{\sqrt{2}P^{+}}P_{11}\bar{D}_{A}
\end{eqnarray}
where $\hat{i}$ is an $SO(8)$ vector index. This object can be written in the compact form
\begin{equation}\label{eq502}
G_{A} = \frac{1}{2P^{+}}P^{m}(\Gamma^{+}\Gamma_{m}D)_{A}
\end{equation}
It will be useful to keep in mind the following $SO(9)$ relations which can be deduced from \eqref{eq501}, \eqref{eq61}:
\begin{equation}\label{eq503}
\begin{aligned}
\{D_{A}, D_{B}\} &= -2\sqrt{2}\,\delta_{AB}P^{-}\hspace{2mm},\\
\{\bar{D}_{A}, \bar{D}_{B}\} &= -2\sqrt{2}\,\delta_{AB}P^{+}\hspace{2mm},
\end{aligned}
\begin{aligned}
\{D_{A}, \bar{D}_{B}\} &= 2[(\gamma^{9}\gamma^{\hat{i}})_{AB}P_{\hat{i}} - \delta_{AB}P_{11}]\\
\{\bar{D}_{A}, D_{B}\} &= 2[-(\gamma^{9}\gamma^{\hat{i}})_{AB}P_{\hat{i}} - \delta_{AB}P_{11}]
\end{aligned}
\end{equation}
where $D_{A}, \bar{D}_{A}$ are given by
\begin{eqnarray}
D_{A} &=& P_{A} + \sqrt{2}i\Theta_{A}P^{-} - i(\gamma^{9}\gamma^{\hat{i}}\bar{\Theta})_{A}P_{\hat{i}} + i\bar{\Theta}_{A}P_{11}\\
\bar{D}_{A} &=& \bar{P}_{A} + \sqrt{2}i\bar{\Theta}_{A}P^{+} + i(\gamma^{9}\gamma^{\hat{i}}\Theta)_{A}P_{\hat{i}} + i\Theta_{A}P_{11}
\end{eqnarray}
or in a more compact form
\begin{eqnarray}
D_{A} &=& P_{A} + \sqrt{2}i\Theta_{A}P^{-} + \Gamma^{I}_{AB}\bar{\Theta}^{B}P_{I}\\
\bar{D}_{A} &=& \bar{P}_{A} + \sqrt{2}i\bar{\Theta}_{A}P^{+} + \Gamma^{I}_{AB}\Theta^{B}P_{I}
\end{eqnarray}
where $\Gamma^{I}_{A\bar{B}} = (-i(\gamma^{9}\gamma^{\hat{i}})_{AB},i\delta_{AB})$, $\Gamma^{I}_{\bar{A}B} = (i(\gamma^{9}\gamma^{\hat{i}})_{AB},i\delta_{AB})$.
 Using the equations \eqref{eq502}, \eqref{eq503} one can show that
\begin{eqnarray}
\{G_{A}, \bar{D}_{B}\} &=& 0\label{eq411}\\
\{G_{A}, G_{B}\} &=& \frac{\sqrt{2}}{P^{+}}(P^{m}P_{m})\delta_{AB}
\end{eqnarray}
Notice that the nilpotency of $\tilde{Q}$ no longer requires the validity of the $SO(9)$ pure spinor constraint $\Lambda^{A}\Gamma^{I}_{AB}\bar{\Lambda}^{B} = 0$ as can be seen from \eqref{eq411}. A further similarity transformation induced by the operator
\begin{equation}
\hat{R} = -\frac{1}{\sqrt{2}P^{+}}(\Theta^{A}\Gamma^{I}_{AB}\bar{P}^{B})P_{I}
\end{equation}
will transform the operators $\bar{D}_{A}$, $G_{A}$ into
\begin{eqnarray}
\hat{\bar{D}}_{A} &=& \bar{P}_{A} + \sqrt{2}i\bar{\Theta}^{A}P^{+}\\
\hat{G}_{A} &=& P_{A} - \frac{i}{\sqrt{2}P^{+}}(P^{m}P_{m})\Theta_{A}
\end{eqnarray}
Hence the pure spinor BRST operator will take the form
\begin{equation}
\tilde{\tilde{Q}} = \Lambda^{A}\hat{G}_{A} + \bar{\Lambda}^{A}\hat{\bar{D}}_{A}
\end{equation}
The supersymmetry invariance of this operator follows from the supersymmetry invariance of $\hat{G}_{A}$ and $\hat{\bar{D}}_{A}$ under the operators
\begin{eqnarray}
\hat{\bar{Q}}_{A} &=& \bar{P}_{A} - \sqrt{2}i\bar{\Theta}_{A}P^{+}\\
\hat{Q}_{A} &=& P_{A} + \frac{i}{\sqrt{2}P^{+}}(P^{m}P_{m})\Theta_{A} - \frac{i}{\sqrt{2}P^{+}}P_{I}\Gamma^{I}_{AD}\hat{\bar{Q}}_{D} 
\end{eqnarray}
which are the $\tilde{R}$-transformed versions of the supersymmetry generators
\begin{eqnarray}
Q_{A} &=& P_{A} - \sqrt{2}iP^{-}\Theta_{A} + i(\gamma^{9}\gamma^{\hat{i}}\bar{\Theta})_{A}P_{\hat{i}} - i\bar{\Theta}_{A}P_{11}\\
\bar{Q}_{A} &=& \bar{P}_{A} - \sqrt{2}iP^{+}\bar{\Theta}_{A} - (\gamma^{9}\gamma^{\hat{i}}\Theta)_{A}P_{\hat{i}} - i\Theta_{A}P_{11}
\end{eqnarray} 
\subsection{Light-cone equations of motion}
The physical fields are contained in the ghost number 3 superfield $V = \Lambda^{\alpha}\Lambda^{\beta}\Lambda^{\sigma}C_{\alpha\beta\sigma}$ \cite{Berkovits:2002uc}. This superfield can be written in $SO(9)$ notation as
\begin{eqnarray}
V &=& \Lambda^{A}\Lambda^{B}\Lambda^{C}C_{(+A)(+B)(+C)} + 3\bar{\Lambda}^{A}\Lambda^{B}\Lambda^{C}C_{(-A)(+B)(+C)}\nonumber\\
&& + 3\bar{\Lambda}^{A}\bar{\Lambda}^{B}\Lambda^{C}C_{(-A)(-B)(+C)} + \bar{\Lambda}^{A}\bar{\Lambda}^{B}\bar{\Lambda}^{C}C_{(-A)(-B)(-C)},\label{eq504}
\end{eqnarray}
where the signs $\pm$ come from the splitting $SO(10,1)\rightarrow SO(1,1)\times SO(9)$. The use of the gauge transformation $\delta V = \tilde{\tilde{Q}}\Omega$, with $\Omega$ being an arbitrary ghost number 2 superfield, allows us to cancel out the last three terms in \eqref{eq504}:
\begin{eqnarray}
\tilde{\tilde{Q}}\Omega &=& \Lambda^{A}\Lambda^{B}\Lambda^{C}\hat{G}_{A}\Omega_{(+B)(+C)} + 2\Lambda^{A}\bar{\Lambda}^{B}\Lambda^{C}\hat{G}_{A}\Omega_{(-B)(+C)} + \Lambda^{A}\bar{\Lambda}^{B}\bar{\Lambda}^{C}\hat{G}_{A}\Omega_{(-B)(-C)}\nonumber\\
&& + \bar{\Lambda}^{A}\Lambda^{B}\Lambda^{C}\hat{\bar{D}}_{A}\Omega_{(+B)(+C)} + 2\bar{\Lambda}^{A}\bar{\Lambda}^{B}\Lambda^{C}\hat{\bar{D}}_{A}\Omega_{(-B)(+C)} + \bar{\Lambda}^{A}\bar{\Lambda}^{B}\bar{\Lambda}^{C}\hat{\bar{D}}_{A}\Omega_{(-B)(-C)},\nonumber
\end{eqnarray}
after conveniently choosing $\Omega_{(-B)(-C)}$, $\Omega_{(+B)(-C)}$, $\Omega_{(+B)(+C)}$. Therefore we are left with 
\begin{equation}\label{eq512}
V = \Lambda^{A}\Lambda^{B}\Lambda^{C}C_{ABC},
\end{equation}
where we have dropped the $SO(1,1)$ index for convenience. The $\tilde{\tilde{Q}}$-closedness condition for $V$ implies the following equations for $C_{BCD}$:
\begin{eqnarray}
\hat{\bar{D}}_{A}C_{BCD} &=& (\Gamma^{J})_{A(B}C_{|J|CD)} + \delta_{(BC}\chi_{D)A}\label{eq510}\\
\hat{G}_{A}C_{BCD} &=& \delta_{(AB}\xi_{CD)} + (\Gamma^{JK})_{A(B}C_{|JK|CD)} + (\Gamma^{JKL})_{A(B}C_{|JKL|CD)},\label{eq511}
\end{eqnarray}
where $\chi_{DA}$, $\xi_{CD}$, $C_{JCD}$, $C_{JKCD}$, $C_{JKLCD}$ are $SO(9)$ p-form-bispinors. Each of these possesses a certain symmetry determined by \eqref{eq510}, \eqref{eq511}. To find the physical spectrum and the corresponding equations of motion, we should solve these equations subject to the constraints:
\begin{eqnarray}
\{\hat{\bar{D}}_{A}, \hat{\bar{D}}_{B}\} &=& -2\sqrt{2}P^{+}\delta_{AB}\label{eq522}\\
\{\hat{G}_{A}, \hat{G}_{B}\} &=& \frac{\sqrt{2}}{P^{+}}(P^{m}P_{m})\delta_{AB}\label{eq521}
\end{eqnarray}

A way to solve this constrained system of equations is the following: Let us choose the only non-zero component of the spinor $\Lambda^{A}$ to be $\Lambda^{+0}$. This choice will imply $\bar{\Lambda}^{-i} = \bar{\Lambda}^{+0} = 0$, where $i$ is the usual $SO(7)$ vector index. With these constraints, the only $\hat{\bar{D}}_{A}$ that act non-trivially on $C_{(+0)(+0)(+0)}$ are $\hat{\bar{D}}_{-i}$ and $\hat{\bar{D}}_{+0}$. Therefore, we will have $2^{8}$ states in $C_{(+0)(+0)(+0)}$: 128 bosonic and 128 fermionic states. The other componens of $C_{ABC}$ can be shown to be related to $C_{(+0)(+0)(+0)}$ by $SO(9)$ rotations (see Appendix \ref{appC}) given by the operator
\begin{equation}
R^{IJ} = \frac{1}{\sqrt{8\sqrt{2}P^{+}}}(\hat{\bar{D}}\Gamma^{IJ}\hat{\bar{D}}),
\end{equation}
which satisfies the algebra
\begin{equation}
[R^{IJ}, R^{KL}] = \eta^{IK}R^{JL} - \eta^{JK}R^{IL} - \eta^{IL}R^{JK} + \eta^{JL}R^{IK}.
\end{equation} 
The 128 fermionic states can be adequately represented by the lowest order term in $\tilde{f}_{JD}$:
\begin{equation}\label{eq530}
C_{BCD} = (\Gamma^{J})_{(BC}\tilde{f}_{|J|D)},
\end{equation}
where $\tilde{f}_{JD}$ is $\Gamma$-traceless. The 128 bosonic states can be accommodated in the $SO(9)$ traceless symmetric tensor $g_{JK}$ and the 3-form $H^{LMN}$. Therefore we can write
\begin{equation}\label{eq531}
C_{JCD} = a(\Gamma^{K})_{CD}g_{JK} + b(\Gamma_{JKLM})_{CD}H^{KLM}
\end{equation}
After replacing \eqref{eq530}, \eqref{eq531} in \eqref{eq510} one obtains
\begin{eqnarray}
\Gamma^{J}_{(BC}\bar{D}_{|A}\tilde{f}_{J|D)} &=& a(\Gamma^{K})_{A(B}(\Gamma^{J})_{CD)}g_{JK} + b(\Gamma_{J})_{A(B}(\Gamma^{JKLM})_{CD)}H_{KLM} + \frac{2b}{3}\delta_{(BC}(\Gamma^{KLM})_{D)A}H_{KLM}\nonumber\\
\end{eqnarray}
Next we use the $SO(9)$ Fierz identities
\begin{eqnarray}
\delta_{(BC}(\Gamma^{KLM})_{D)A} &=& 3(\Gamma^{[K})_{(BC}(\Gamma^{LM]})_{D)A} + (\Gamma_{J})_{(BC}(\Gamma^{JKLM})_{D)A},\\
(\Gamma^{JKLM})_{(BC}(\Gamma_{J})_{D)A} &=& -(\Gamma_{J})_{(BC}(\Gamma^{JKLM})_{D)A},
\end{eqnarray}
which can be found by using the Mathematica package GAMMA \cite{Gran:2001yh}, to obtain
\begin{eqnarray}
\Gamma^{J}_{(BC}\bar{D}_{|A}\tilde{f}_{J|D)} &=& a(\Gamma^{J})_{(BC}(\Gamma^{K})_{D)A}g_{JK} + 2b(\Gamma^{J})_{(BC}(\Gamma^{LM})_{D)A}H_{JLM} - \frac{b}{3}(\Gamma_{J})_{(BC}(\Gamma^{JKLM})_{D)A}H_{KLM},\nonumber\\
\end{eqnarray}
which implies
\begin{equation}
\hat{\bar{D}}_{A}\tilde{f}_{JD} = a(\Gamma^{K})_{AD}g_{JK} - 2b(\Gamma^{LM})_{AD}H_{JLM} - \frac{b}{3}(\Gamma_{JKLM})_{AD}H^{KLM},
\end{equation}
where the constants $a$, $b$ will be determined from supersymmetry. To do this we should know how $\hat{\bar{D}}_{A}$ acts on $g_{JK}$ and $H_{KLM}$. An educated guess based on linearity and symmetry properties is
\begin{eqnarray}
\hat{\bar{D}}_{A}g_{JK} &=& -2\sqrt{2}P^{+}[(\Gamma_{J})_{AE}\tilde{f}_{K\,E} + (\Gamma_{K})_{AE}\tilde{f}_{J\,E}],\\
\hat{\bar{D}}_{A}H^{KLM} &=& -2\sqrt{2}P^{+}[(\Gamma^{KL})_{AE}\tilde{f}^{M}_{E} - (\Gamma^{KM})_{AE}\tilde{f}^{L}_{E} + (\Gamma^{LM})_{AE}\tilde{f}^{K}_{E}],
\end{eqnarray}
where the factor $-2\sqrt{2}P^{+}$ was chosen for convenience. These equations of motion should satisfy the supersymmetry algebra \eqref{eq522}. This requirement fixes the values of $a$, $b$ to be $a=\frac{1}{4}$, $b=\frac{1}{72}$. Therefore the whole set of light-cone equations of motion is
\begin{eqnarray}
\hat{\bar{D}}_{A}g_{JK} &=& -2\sqrt{2}P^{+}[(\Gamma_{J})_{AE}\tilde{f}_{K\,E} + (\Gamma_{K})_{AE}\tilde{f}_{J\,E}],\\
\hat{\bar{D}}_{A}\tilde{f}_{JD} &=& \frac{1}{4}(\Gamma^{K})_{AD}g_{JK} + \frac{1}{72}[(\Gamma_{JKLM})_{AD} + 6\eta^{JK}(\Gamma^{LM})_{AD}]H^{KLM},\\
\hat{\bar{D}}_{A}H^{KLM} &=& -2\sqrt{2}P^{+}[(\Gamma^{KL})_{AE}\tilde{f}^{M}_{E} - (\Gamma^{KM})_{AE}\tilde{f}^{L}_{E} + (\Gamma^{LM})_{AE}\tilde{f}^{K}_{E}].
\end{eqnarray}
These expressions are the same equations of motion obtained for $D=11$ linearized supergravity from the light-cone $D=11$ Brink-Schwarz-like superparticle \cite{Green:1999by}. 

\vspace{2mm}
The mass-shell condition can be obtained from \eqref{eq521} after using the tracelessness condition for $C_{BCD}$, which is necessary to have a non-trivial vertex operator $V$. This condition gives rise to the equation:
\begin{equation}
\hat{G}_{A}C_{BCD} + \hat{G}_{B}C_{ACD} + \hat{G}_{C}C_{ABD} + \hat{G}_{D}C_{ABC} = 0
\end{equation}
which has solution only if $\hat{G}_{A}C_{BCD} = 0$. This result, together with \eqref{eq521}, implies that $k^{m}k_{m} = 0$,  where $k^{m}$ is the momentum. Consequently, $C_{BCD}$ depends only on $\bar{\Theta}$, $C_{BCD}=C_{BCD}(\bar{\Theta})$. To obtain the pure spinor vertex operator in the $Q$-cohomology one just performs the similarity transformation generated by $-(R+\hat{R})$. The result is
\begin{eqnarray}
V &=& V(\hat{\bar{\Theta}})e^{ik.X}
\end{eqnarray}
where $\hat{{\bar{\Theta}}}^{A} = \bar{\Theta}^{A} - \frac{i}{\sqrt{2}P^{+}}\Theta_{B}(\Gamma^{I})^{AB}k_{I}$.
\section{Remarks}
The equivalence of cohomologies for the $D=11$ Brink-Schwarz-like superparticle and the $D=11$ pure spinor superparticle is strong evidence that the two models describe the same physical theory. Our method to demonstrate the equivalence uses ideas that were applied previously to the $D=10$ case (e.g., the group decomposition $SO(10,1)\rightarrow SO(1,1)\times SO(9)$), and introduces a parametrization of $D=11$ objects (the group decomposition $SO(10,1)\rightarrow SO(3,1)\times SO(7)$) which was useful for analyzing the light-cone pure spinor cohomology.

\vspace{2mm}
The equations of motion in $D=9$ superspace found in this paper, by studying the light-cone pure spinor cohomology, match the light-cone equations of motion presented in \cite{Green:1999by}. We conclude that the $D=11$ pure spinor superparticle is a good model to study $D=11$ linearized supergravity in a manifestly covariant way.

\subsection*{Acknowledgements}
I would like to thank Nathan Berkovits for very useful discussions, and FAPESP grant 15/23732-2 for financial support.

\begin{appendices}
\section{$\Gamma$-matrices of $SO(10,1)$}\label{appA}

We will denote $SO(10,1)$ vector indices by $m, n, \ldots$ and $SO(9,1)$ vector indices by $\hat{m}, \hat{n}, \ldots$. In addition, we will denote $SO(10,1)$ spinor indices by $\alpha, \beta, \ldots$ and $SO(9,1)$ spinor indices by $\mu, \nu, \ldots$. As usual, we add a new matrix, $\Gamma^{10}$, to the set of $SO(9,1)$ gamma matrices $\{\Gamma^{\hat{m}}\}$, which is numerically equal to the chirality matrix $\Gamma^{(9,1)}$ in $D = (9,1)$:
\begin{equation}
\Gamma^{10} = \Gamma^{(9,1)} = \begin{pmatrix}
I_{16\times 16} & 0\\
0 & -I_{16\times 16}
\end{pmatrix}
\end{equation}
This matrix satisfies the properties $\{\Gamma^{m}, \Gamma^{10}\} = 0$, for $m = 0, \ldots , 9$, and $(\Gamma^{10})^{2} = 1$. The chirality matrix $\Gamma$ in $D = (10,1)$ is given by:
\begin{equation}
\Gamma = \Gamma^{0}\Gamma^{1}\ldots \Gamma^{9}\Gamma^{10} = \Gamma^{(9,1)}\Gamma^{10} = (\Gamma^{10})^{2} = 1
\end{equation}
which reflects the fact that we don't have Weyl (anti-Weyl) spinors in this case. However, we can have Majorana spinors.
It is easy to see that $C = \Gamma^{0}$ satisfies the definition of the charge conjugation matrix\footnote{We know that for $D=(9,1)$, $C^{(9,1)} = \Gamma^{0}$ is the charge conjugation matrix, so we just need to show that $C = \Gamma^{0}$ obeys $C\Gamma^{10} = -(\Gamma^{10})^{T}C$, which is trivial since $\Gamma^{10}$ is symmetric and $\{\Gamma^{10}, \Gamma^{0}\} = 0$.} $C\Gamma^{m} = -(\Gamma^{m})^{T}C$.\\
For two Majorana spinors $\Theta$ and $\Psi$, we have $\bar{\Theta}\Gamma^{m}\Psi = \Theta^{T} C\Gamma^{m}\Psi$. This result can be viewed in terms of $SO(9,1)$ components:
\begin{equation}
\Theta^{T} C\Gamma^{m}\Psi =  \begin{pmatrix} \Theta^{\mu} & \Theta_{\mu}
\end{pmatrix} \begin{pmatrix}
\gamma^{\hat{m}}_{\mu\nu} & 0\\
0 & -(\gamma^{\hat{m}})^{\mu\nu}
\end{pmatrix} \begin{pmatrix}
\Psi^{\nu} & \Psi_{\nu}
\end{pmatrix},
\end{equation}
\begin{equation}
\Theta^{T} C\Gamma^{10}\Psi =  \begin{pmatrix} \Theta^{\mu} & \Theta_{\mu}
\end{pmatrix} \begin{pmatrix}
0 & -1\\
-1 & 0
\end{pmatrix} \begin{pmatrix}
\Psi^{\nu} & \Psi_{\nu}
\end{pmatrix},
\end{equation}
where $m = 0, \ldots, 9$ and $\gamma^{\hat{m}}_{\mu\nu}$, and $(\gamma^{\hat{m}})^{\mu\nu}$ are the $SO(9,1)$ $\gamma$-matrices. 
It is useful to mention that the index structure of the charge conjugation matrix is $C_{\alpha\beta}$. So, the $\Gamma$-matrices have index structure $(\Gamma^{m})^{\alpha}_{\hspace{2mm}\beta}$ and when are multiplied by the charge conjugation matrix (or its inverse) we obtain the corresponding matrices $(\Gamma^{m})_{\alpha\beta}$ and $(\Gamma^{m})^{\alpha\beta}$.

\vspace{2mm}
Next we will show explicitly the form of the gamma matrices. For $D=(9,1)$, we have:

\begin{eqnarray}
\begin{aligned}
(\gamma^{0})^{\alpha\beta} &= \begin{pmatrix} 1_{8\times 8} & 0\\
0 & 1_{8\times 8}
\end{pmatrix} \hspace{8mm},\\
(\gamma^{9})^{\alpha\beta} &= \begin{pmatrix}1_{8\times 8} & 0\\
0 & -1_{8\times 8}
\end{pmatrix}\hspace{5.2mm},\\
(\gamma^{\hat{i}})^{\alpha\beta} &= \begin{pmatrix} 0 & \sigma^{\hat{i}}_{a\dot{a}}\\
\sigma^{\hat{i}}_{\dot{b}b} & 0\end{pmatrix}\hspace{12.6mm},\\
(\gamma^{+})^{\alpha\beta} &= \begin{pmatrix}\sqrt{2}_{8\times 8} & 0\\
0 & 0
\end{pmatrix}\hspace{10.5mm},\\
(\gamma^{-})^{\alpha\beta} &= \begin{pmatrix}0 & 0\\
0 & \sqrt{2}_{8\times 8}
\end{pmatrix}\hspace{10.5mm},\\
\end{aligned}\hspace{4mm}
\begin{aligned}
 (\gamma^{0})_{\alpha\beta} &= \begin{pmatrix} -1_{8\times 8} & 0\\
0 & -1_{8\times 8}
\end{pmatrix}\\
 (\gamma^{9})_{\alpha\beta} &= \begin{pmatrix} 1_{8\times 8} & 0\\
0 & -1_{8\times 8}
\end{pmatrix}\\
 (\gamma^{\hat{i}})_{\alpha\beta} &= \begin{pmatrix}0 & \sigma^{\hat{i}}_{a\dot{a}}\\
\sigma^{\hat{i}}_{\dot{b}b} & 0
\end{pmatrix}\\
 (\gamma^{+})_{\alpha\beta} &= \begin{pmatrix}0 & 0\\
0 & -\sqrt{2}_{8\times 8}
\end{pmatrix}\\
 (\gamma^{-})_{\alpha\beta} &= \begin{pmatrix}-\sqrt{2}_{8\times 8} & 0\\
0 & 0
\end{pmatrix},
\end{aligned}
\end{eqnarray}
where each entry is an $8\times 8$ matrix and $\hat{i}$ is a $SO(8)$ vector index. The matrices $\gamma^{\pm}$ are defined by
\begin{equation*}
\gamma^{\pm} = \frac{1}{\sqrt{2}}(\gamma^{0} \pm \gamma^{9})
\end{equation*}
The $\sigma^{\hat{i}}$ matrices are defined by

$$
\begin{array}{cl}
\sigma^{1}_{a\dot{a}} = \epsilon\otimes\epsilon\otimes\epsilon \hspace{10mm} & \sigma^{5}_{a\dot{a}} = \tau^{3}\otimes\epsilon\otimes 1\\
\sigma^{2}_{a\dot{a}} = 1\otimes\tau^{1}\otimes\epsilon \hspace{10mm} & \sigma^{6}_{a\dot{a}} = \epsilon\otimes 1\otimes\tau^{1}\\
\sigma^{3}_{a\dot{a}} = 1\otimes\tau^{3}\otimes\epsilon \hspace{10mm} & \sigma^{7}_{a\dot{a}} = \epsilon\otimes 1\otimes\tau^{3}\\
\sigma^{4}_{a\dot{a}} = \tau^{1}\otimes\epsilon\otimes 1 \hspace{10mm} & \sigma^{8}_{a\dot{a}} = 1\otimes 1\otimes 1
\end{array}
$$
where $\epsilon = i\tau^{2}$ and $\tau^{1}$, $\tau^{2}$, $\tau^{3}$ are the usual Pauli matrices. The $\sigma^{\hat{i}}_{\dot{a}a}$ are symmetric ($\sigma^{\hat{i}}_{\dot{a}a} = (\sigma^{\hat{i}}_{a\dot{a}})^{T}$) and satisfy the following relations:
\begin{eqnarray*}
\sigma^{\hat{i}}_{a\dot{a}}\sigma^{\hat{j}}_{\dot{a}b} + \sigma^{\hat{j}}_{a\dot{a}}\sigma^{\hat{i}}_{\dot{a}b} &=& 2\delta^{\hat{i}\hat{j}}\delta_{ab}\\
\sigma^{\hat{i}}_{\dot{a}a}\sigma^{\hat{j}}_{a\dot{b}} + \sigma^{\hat{j}}_{\dot{a}a}\sigma^{\hat{i}}_{a\dot{b}} &=& 2\delta^{\hat{i}\hat{j}}\delta_{\dot{a}\dot{b}}\\
\sigma^{\hat{i}}_{\dot{a}b}\sigma^{\hat{i}}_{a\dot{c}} + \sigma^{\hat{i}}_{\dot{a}a}\sigma^{\hat{i}}_{b\dot{c}} &=& 2\delta_{ab}\delta_{\dot{a}\dot{c}} 
\end{eqnarray*}

Similarly, for $D=(10,1)$, we have:
\begin{eqnarray}
\sbox0{$\begin{matrix}0&0\\0&-\sqrt{2}i\end{matrix}$}
\begin{aligned}
(\Gamma^{\hat{i}})^{\alpha\beta} &= \begin{pmatrix}
-i\gamma^{\hat{i}\,AB} & \makebox[\wd0]{\large $O$}\\
\makebox[\wd0]{\large $O$} & i\gamma^{\hat{i}}_{AB}
\end{pmatrix}\hspace{14mm},\\
(\Gamma^{11})^{\alpha\beta} &= \begin{pmatrix}
\makebox[\wd0]{\large $O$} & -i\\
-i & \makebox[\wd0]{\large $O$}
\end{pmatrix}\hspace{14mm},\\
(\Gamma^{+})^{\alpha\beta} &= \begin{pmatrix}
\begin{pmatrix}
-\sqrt{2}i & 0\\ 
0 & 0
\end{pmatrix} & \makebox[\wd0]{\large $O$}\\
\makebox[\wd0]{\large $O$} & \begin{pmatrix}
0 & 0\\
0 & -\sqrt{2}i
\end{pmatrix}
\end{pmatrix}\hspace{2mm},\\
(\Gamma^{-})^{\alpha\beta} &= \begin{pmatrix}
\begin{pmatrix}
0 & 0\\ 
0 & -\sqrt{2}i
\end{pmatrix} & \makebox[\wd0]{\large $O$}\\
\makebox[\wd0]{\large $O$} & \begin{pmatrix}
-\sqrt{2}i & 0\\
0 & 0
\end{pmatrix}
\end{pmatrix}\hspace{2mm},
\end{aligned}
\begin{aligned}
(\Gamma^{\hat{i}})_{\alpha\beta} &= \begin{pmatrix}
i\gamma^{\hat{i}}_{AB} & \makebox[\wd0]{\large $O$}\\
\makebox[\wd0]{\large $O$} & -i\gamma^{\hat{i}\,AB}
\end{pmatrix}\\
(\Gamma^{11})_{\alpha\beta} &= \begin{pmatrix}
\makebox[\wd0]{\large $O$} & i\\
i & \makebox[\wd0]{\large $O$}
\end{pmatrix}\\
(\Gamma^{+})_{\alpha\beta} &= \begin{pmatrix}
\begin{pmatrix}
0 & 0\\ 
0 & -\sqrt{2}i
\end{pmatrix} & \makebox[\wd0]{\large $O$}\\
\makebox[\wd0]{\large $O$} & \begin{pmatrix}
-\sqrt{2}i & 0\\
0 & 0
\end{pmatrix}
\end{pmatrix}\\
(\Gamma^{-})_{\alpha\beta} &= \begin{pmatrix}
\begin{pmatrix}
-\sqrt{2}i & 0\\ 
0 & 0
\end{pmatrix} & \makebox[\wd0]{\large $O$}\\
\makebox[\wd0]{\large $O$} & \begin{pmatrix}
0 & 0\\
0 & -\sqrt{2}i
\end{pmatrix}
\end{pmatrix}
\end{aligned}
\end{eqnarray}
where $A$, $B$ are $SO(9)$ spinor indices. Notice that each $\Gamma$ matrix is $32\times 32$.

\vspace{2mm}
To construct the above representation of the $\Gamma$ matrices, we used a basis convenient for dealing with $SO(8)$ objects. Hence, an arbitrary $D=11$ spinor $\chi^{\alpha}$ is written in this basis as
\begin{equation}
\chi^{\alpha} = \begin{pmatrix}
\chi^{a}\\
\chi^{\dot{a}}\\
\bar{\chi}^{a}\\
\bar{\chi}^{\dot{a}}
\end{pmatrix}
\end{equation}
This was the convention used in \eqref{eq500}. This is useful when $SO(8)$ objects are needed, as in Section 3. However, when analyzing the light-cone structure of the pure spinor cohomology and vertex operators, we need to deal with $SO(9)$ objects. So, we define the following change of basis matrix:
\begin{equation}
M_{cbm} = \left(\begin{matrix}
0 & 0 & 1 & 0\\
0 & 1 & 0 & 0\\
1 & 0 & 0 & 0\\
0 & 0 & 0 & 1
\end{matrix}\right)
\end{equation}
where each entry represents an $8\times 8$ matrix. Using this matrix we find the corresponding $\Gamma$ matrices in this new basis:
\begin{eqnarray}
\sbox0{$\begin{matrix}0&0\\0&-\sqrt{2}i\end{matrix}$}
\begin{aligned}
(\Gamma^{\hat{i}})^{\alpha\beta} &= i\begin{pmatrix}
\makebox[\wd0]{\large $O$} & (\gamma^{9}\gamma^{\hat{i}})^{AB}\\
-(\gamma^{9}\gamma^{\hat{i}})_{AB} & \makebox[\wd0]{\large $O$}
\end{pmatrix}\hspace{2mm},\\
(\Gamma^{11})^{\alpha\beta} &= -i\begin{pmatrix}
 \makebox[\wd0]{\large $O$} &  I^{AB}\\
I_{AB} &  \makebox[\wd0]{\large $O$}
\end{pmatrix}\hspace{6.5mm},\\
(\Gamma^{+})^{\alpha\beta} &= \left(\begin{array}{cc}
\makebox[\wd0]{$O$} & \makebox[\wd0]{$O$}\\
\makebox[\wd0]{$O$} & -\sqrt{2}i
\end{array}\right)\hspace{9.8mm},\\
(\Gamma^{-})^{\alpha\beta} &= \left(\begin{array}{cc}
-\sqrt{2}i & \makebox[\wd0]{$O$}\\
\makebox[\wd0]{$O$} & \makebox[\wd0]{$O$}
\end{array}\right)\hspace{9.8mm},\\
\end{aligned}
\begin{aligned}
(\Gamma^{\hat{i}})_{\alpha\beta} &= i\begin{pmatrix}
 \makebox[\wd0]{\large $O$} & -(\gamma^{9}\gamma^{\hat{i}})^{AB}\\
(\gamma^{9}\gamma^{\hat{i}})_{AB} &  \makebox[\wd0]{\large $O$}
\end{pmatrix}\\
(\Gamma^{11})_{\alpha\beta} &= i\begin{pmatrix}
\makebox[\wd0]{$O$} & I_{AB}\\
I^{AB} & \makebox[\wd0]{$O$}
\end{pmatrix}\\
(\Gamma^{+})_{\alpha\beta} &= 
\left(\begin{array}{cc}
-\sqrt{2}i & \makebox[\wd0]{$O$}\\
\makebox[\wd0]{$O$} & \makebox[\wd0]{$O$}
\end{array}\right)\\
(\Gamma^{-})_{\alpha\beta} &= \left(\begin{array}{cc}
\makebox[\wd0]{$O$} & \makebox[\wd0]{$O$}\\
\makebox[\wd0]{$O$} & -\sqrt{2}i 
\end{array}\right)
\end{aligned}
\end{eqnarray}
where $I_{AB}$ is the $SO(9)$ identity matrix, $A$, $B$ are $SO(9)$ spinor indices, and $\hat{i} = 1,\ldots,8$. Each entry in the above matrices is $16\times 16$.

\section{$SO(10,1) \rightarrow SO(3,1)\times SO(7)$}\label{appB}
Here we will explain the $\pm$ notation, and construct explicitly a different representation for the $SO(10,1)$ gamma matrices. Let us define the \emph{raising} and \emph{lowering} $\Gamma$-matrices:
\begin{eqnarray}
\Gamma^{\pm 0+1} &=& \frac{1}{2}(\pm\Gamma^{0} + \Gamma^{1})\\
\Gamma^{2\pm 3i} &=& \frac{1}{2}(\Gamma^{2} \pm i\Gamma^{3})
\end{eqnarray}
These $\Gamma^{\pm}$ matrices act on an arbitrary spinor $\chi$ as follows: 
\begin{eqnarray}
\begin{aligned}
\Gamma^{0+1}|- + a> &= |+ + a>\hspace{1mm},\\
\Gamma^{0+1}|- - a> &= |+ - a>\hspace{1mm},\\
\Gamma^{-0+1}|+ + a> &= |- + a> \hspace{1mm},\\
\Gamma^{-0+1}|+ - a> &= |- - a>\hspace{1mm},\\
\Gamma^{j}|- - 0> &= |- - j>,
\end{aligned}\hspace{1mm}
&\begin{aligned}
\Gamma^{3+4i}|- - a> &= -|- + a>\hspace{3mm},\\
\Gamma^{3+4i}|+ - a> &= |+ + a>\hspace{6mm},\\
\Gamma^{3-4i}|- + a> &= -|- + a>\hspace{3mm},\\
\Gamma^{3-4i}|+ + a> &= |+ - a>\hspace{6mm},\\
\Gamma^{j}|- - j> &= |- - 0>\hspace{1.8mm},
\end{aligned}\hspace{1mm}\nonumber
&
\begin{aligned}
\Gamma^{j}|+ + 0> &= |+ + j> \\
\Gamma^{j}|+ + j> &= |+ + 0> \\
\Gamma^{j}|- + 0> &= -|- + j> \\
\Gamma^{j}|- + j> &= -|- + 0>\\
\Gamma^{j}|+ - 0> &= -|+ - j>
\end{aligned}\\
&
\begin{aligned}
\hspace{2mm}\Gamma^{j}|+ - j> &= -|+ - 0>\nonumber
\end{aligned}& \\
\end{eqnarray}
and any other relation vanishes. In these formulae we have made the identification $|\pm\pm a> = \chi^{\pm\pm a}$ with $a = 0, i$. It is clear that these relations are consistent with the $SO(10,1)$ Clifford algebra. With these rules, one can construct the respective representation:
\begin{equation}
\begin{aligned}
(\Gamma^{0+1})^{\alpha}_{\hspace{2mm}\beta} &= \begin{pmatrix}
0 & 0 & 0 & 1\\
0 & 0 & 0 & 0\\
0 & 1 & 0 & 0\\
0 & 0 & 0 & 0
\end{pmatrix}\hspace{2mm},
\end{aligned}\hspace{2mm}
\begin{aligned}
(\Gamma^{-0+1})^{\alpha}_{\hspace{2mm}\beta} &= \begin{pmatrix}
0 & 0 & 0 & 0\\
0 & 0 & 1 & 0\\
0 & 0 & 0 & 0\\
1 & 0 & 0 & 0
\end{pmatrix}
\end{aligned}
\end{equation}
Here and throughout this Appendix, each entry will represent an $8\times 8$ matrix unless otherwise stated. Now, it is easy to calculate the explicit form of the matrices $(\Gamma^{0})^{\alpha}_{\hspace{2mm}\beta}$, $(\Gamma^{1})^{\alpha}_{\hspace{2mm}\beta}$:
\begin{equation}
\begin{aligned}
(\Gamma^{0})^{\alpha}_{\hspace{2mm}\beta} &= \begin{pmatrix}
0 & 0 & 0 & 1\\
0 & 0 & -1 & 0\\
0 & 1 & 0 & 0\\
-1 & 0 & 0 & 0
\end{pmatrix}\hspace{2mm}, 
\end{aligned}\hspace{2mm}
\begin{aligned}
(\Gamma^{1})^{\alpha}_{\hspace{2mm}\beta} &= \begin{pmatrix}
0 & 0 & 0 & 1\\
0 & 0 & 1 & 0\\
0 & 1 & 0 & 0\\
1 & 0 & 0 & 0
\end{pmatrix}
\end{aligned}
\end{equation}
Similarly, we find
\begin{equation}
\begin{aligned}
(\Gamma^{2+3i})^{\alpha}_{\hspace{2mm}\beta} &= \begin{pmatrix}
0 & 0 & 1 & 0\\
0 & 0 & 0 & 0\\
0 & 0 & 0 & 0\\
0 & -1 & 0 & 0
\end{pmatrix}\hspace{5mm},\\
(\Gamma^{2})^{\alpha}_{\hspace{2mm}\beta} &= \begin{pmatrix}
0 & 0 & 1 & 0\\
0 & 0 & 0 & -1\\
1 & 0 & 0 & 0\\
0 & -1 & 0 & 0
\end{pmatrix}\hspace{2.8mm}, 
\end{aligned}\hspace{2mm}
\begin{aligned}
(\Gamma^{2-3i})^{\alpha}_{\hspace{2mm}\beta} &= \begin{pmatrix}
0 & 0 & 0 & 0\\
0 & 0 & 0 & -1\\
1 & 0 & 0 & 0\\
0 & 0 & 0 & 0
\end{pmatrix}\\
(\Gamma^{3})^{\alpha}_{\hspace{2mm}\beta} &= \begin{pmatrix}
0 & 0 & -i & 0\\
0 & 0 & 0 & -i\\
i & 0 & 0 & 0\\
0 & i & 0 & 0
\end{pmatrix}
\end{aligned}
\end{equation}
However, as already mentioned, there exists an antisymmetric metric tensor $C_{\alpha\beta}$ in $D=11$ dimensions which raises and lower indices. Let us define it as follows:
\begin{equation}
\begin{aligned}
C_{\alpha\beta} = \begin{pmatrix}
0 & -B & 0 & 0\\
B & 0 & 0 & 0\\
0 & 0 & 0 & -B\\
0 & 0 & B & 0
\end{pmatrix}\hspace{2mm},
\end{aligned}\hspace{2mm}
\begin{aligned}
(C^{-1})^{\alpha\beta} = \begin{pmatrix}
0 & B & 0 & 0\\
-B & 0 & 0 & 0\\
0 & 0 & 0 & B\\
0 & 0 & -B & 0
\end{pmatrix}
\end{aligned}
\end{equation}
where $B$ is a diagonal matrix with elements $B_{00}=1$, $B_{jj}=-1$. To preserve the original Clifford algebra we need to multiply the matrices $(\Gamma^{a})^{\alpha}_{\hspace{2mm}\beta}$ by $i$. Now we can find the matrices $(\Gamma^{m})_{\alpha\beta}$, $(\Gamma^{m})^{\alpha\beta}$:
\begin{equation}
\begin{aligned}
\Gamma^{-0+1}_{\alpha\beta} &=& \begin{pmatrix}
0 & 0 & -iB& 0 \\
0 & 0 & 0 & 0\\
-iB & 0 & 0 & 0\\
0 & 0 & 0 & 0
\end{pmatrix}\hspace{2mm},
\end{aligned}\hspace{2mm}
\begin{aligned}
\Gamma^{0+1}_{\alpha\beta} &=& \begin{pmatrix}
0 & 0 & 0 & 0 \\
0 & 0 & 0 & iB\\
0 & 0 & 0 & 0\\
0 & iB & 0 & 0
\end{pmatrix}
\end{aligned}
\end{equation}
and so
\begin{equation}
\begin{aligned}
\Gamma^{0}_{\alpha\beta} &=& \begin{pmatrix}
0 & 0 & iB & 0 \\
0 & 0 & 0 & iB\\
iB & 0 & 0 & 0\\
0 & iB & 0 & 0
\end{pmatrix}\hspace{2mm},
\end{aligned}\hspace{2mm}
\begin{aligned}
\Gamma^{1}_{\alpha\beta} &=& \begin{pmatrix}
0 & 0 & -iB & 0 \\
0 & 0 & 0 & iB\\
-iB & 0 & 0 & 0\\
0 & iB & 0 & 0
\end{pmatrix}
\end{aligned}
\end{equation}
By using $(\Gamma^{m})^{\alpha\beta} = C^{\alpha\delta}C^{\beta\lambda}(\Gamma^{m})_{\delta\lambda}$, we find the matrices $\Gamma^{0\,\alpha\beta}$, $\Gamma^{1\,\alpha\beta}$: 
\begin{equation}
\Gamma^{0\,\alpha\beta} = \begin{pmatrix}
0 & B & 0 & 0\\
-B & 0 & 0 & 0\\
0 & 0 & 0 & B\\
0 & 0 & -B & 0
\end{pmatrix}
\begin{pmatrix}
0 & 0 & iB & 0\\
0 & 0 & 0 & iB\\
iB & 0 & 0 & 0\\
0 & iB & 0 & 0
\end{pmatrix}
\begin{pmatrix}
0 & -B & 0 & 0\\
B & 0 & 0 & 0\\
0 & 0 & 0 & -B\\
0 & 0 & B & 0
\end{pmatrix} 
= \begin{pmatrix}
0 & 0 & iB & 0\\
0 & 0 & 0 & iB\\
iB & 0 & 0 & 0\\
0 & iB & 0 & 0
\end{pmatrix} 
\end{equation}
\begin{equation}
\Gamma^{1\,\alpha\beta} = \begin{pmatrix}
0 & 0 & iB & 0\\
0 & 0 & 0 & -iB\\
iB & 0 & 0 & 0\\
0 & -iB & 0 & 0
\end{pmatrix}
\end{equation}
Analogously, we can find the remaining matrices,
\begin{equation}
\begin{aligned}
(\Gamma^{2+3i})_{\alpha\beta} &= \begin{pmatrix}
0 & 0 & 0 & 0\\
0 & 0 & iB & 0\\
0 & iB & 0 & 0\\
0 & 0 & 0 & 0
\end{pmatrix}\hspace{25mm},\\
(\Gamma^{2})_{\alpha\beta} &= \begin{pmatrix}
0 & 0 & 0 & iB\\
0 & 0 & iB & 0\\
0 & iB & 0 & 0\\
iB & 0 & 0 & 0
\end{pmatrix}\hspace{20.5mm},\\
(\Gamma^{2})^{\alpha\beta} &= \begin{pmatrix}
0 & 0 & 0 & -iB\\
0 & 0 & -iB & 0\\
0 & -iB & 0 & 0\\
-iB & 0 & 0 & 0
\end{pmatrix}\hspace{9.3mm},\\
(\Gamma^{i})_{\alpha\beta} &= \begin{pmatrix}
0 & -iBA & 0 & 0\\
iBA & 0 & 0 & 0\\
0 & 0 & 0 & iBA\\
0 & 0 & -iBA & 0
\end{pmatrix}\hspace{4.5mm}, 
\end{aligned}\hspace{2mm}
\begin{aligned}
(\Gamma^{2-3i})_{\alpha\beta} &= \begin{pmatrix}
0 & 0 & 0 & iB\\
0 & 0 & 0 & 0\\
0 & 0 & 0 & 0\\
iB & 0 & 0 & 0
\end{pmatrix}\\
(\Gamma^{3})_{\alpha\beta} &= \begin{pmatrix}
0 & 0 & 0 & -B\\
0 & 0 & B & 0\\
0 & B & 0 & 0\\
-B & 0 & 0 & 0
\end{pmatrix}\\
(\Gamma^{3})^{\alpha\beta} &= \begin{pmatrix}
0 & 0 & 0 & -B\\
0 & 0 & B & 0\\
0 & B & 0 & 0\\
-B & 0 & 0 & 0
\end{pmatrix}\\
(\Gamma^{i})^{\alpha\beta} &= \begin{pmatrix}
0 & iBA & 0 & 0\\
-iBA & 0 & 0 & 0\\
0 & 0 & 0 & -iBA\\
0 & 0 & iBA & 0
\end{pmatrix}
\end{aligned}
\end{equation}
where $A$ is an $8\times 8$ matrix with non-vanishing elements $A_{0j}=A_{j0}=1$. All these matrices are symmetric and satisfy the desired property: $\Gamma^{m}_{\alpha\beta}\Gamma^{n\,\beta\lambda} + \Gamma^{n}_{\alpha\beta}\Gamma^{m\,\beta\lambda} = 2\eta^{mn}\delta^{\lambda}_{\alpha} $. 

\vspace{2mm}
Finally, the product of two spinors $\chi^{\alpha}\rho_{\alpha}$ will be defined as follows:
\begin{eqnarray*}
\chi^{\alpha}C_{\alpha\beta}\rho^{\beta} &=& -\chi^{++0}\rho^{--0} + \chi^{++i}\rho^{--i} + \chi^{--0}\rho^{++0} - \chi^{--i}\rho^{++i} \\&&- \chi^{+-0}\rho^{-+0} + \chi^{+-i}\rho^{-+i} + \chi^{-+0}\rho^{+-0} - \chi^{-+i}\rho^{+-i}
\end{eqnarray*}
\section{Octonions and $SO(7)$ rotations}\label{appC}
In this Appendix we will show that any component of $C_{BCD}$ can be obtained from $C_{(+0)(+0)(+0)}$ by $SO(9)$ rotations. These rotations are defined by the operator 
\begin{equation}
R^{IJ} = \frac{1}{\sqrt{8\sqrt{2}P^{+}}}\hat{\bar{D}}\Gamma^{IJ}\hat{\bar{D}},
\end{equation}
which satisfy the algebra
\begin{equation}
[R^{IJ}, R^{KL}] = \eta^{IK}R^{JL} - \eta^{JK}R^{IL} - \eta^{IL}R^{JK} + \eta^{JL}R^{IK}
\end{equation}
Therefore, we can use this operator to rotate the \emph{ground state} $C_{(+0)(+0)(+0)}$. To do this let us first write the transformation rule for a general $C_{BCD}$ being acted on by $R^{IJ}$:
\begin{equation}
R^{IJ}C_{BCD} = \frac{1}{\sqrt{2}}(\Gamma^{IJ})_{B}^{\hspace{2mm}E}C_{ECD} + \frac{1}{\sqrt{2}}(\Gamma^{IJ})_{C}^{\hspace{2mm}E}C_{BED} + \frac{1}{\sqrt{2}}(\Gamma^{IJ})_{D}^{\hspace{2mm}E}C_{BCE}
\end{equation}
As explained above, only $\hat{\bar{D}}_{-i}$ and $\hat{\bar{D}}_{+0}$ will act non-trivially on $C_{(+0)(+0)(+0)}$. Thus, we have
\begin{equation}\label{eq600}
(\Gamma^{ij})_{(+k)(-0)}\hat{\bar{D}}_{-k}\hat{\bar{D}}_{+0}C_{(+0)(+0)(+0)} \propto (\Gamma^{ij})_{(+0)}^{\hspace{6mm}E}C_{E(+0)(+0)}
\end{equation}
To solve this equation we recall the notion of octonions \cite{Baez:2001dm}. 

\vspace{2mm}
The octonion mutiplication table can be written in the form
\begin{equation}
e_{i}e_{j} = -\delta_{ij} + \epsilon_{ijk}e_{k}
\end{equation}
which is equivalent to
\begin{equation}
e_{i}e_{j} = \delta_{ij} - i\epsilon_{ijk}e_{k}
\end{equation}
where $\epsilon_{ijk}$ is a totally antisymmetric tensor with value +1 when $(ijk) =$ $(123)$, $(145)$, $(176)$, $(246)$, $(257)$, $(347)$, $(365)$. Now we can identify these octonions as the gamma matrices of the $SO(7)$ Clifford algebra:
\begin{equation}
\Gamma^{i}\Gamma^{j} = \delta^{ij} - i\epsilon^{ijk}\Gamma^{k}
\end{equation}
This equation can be thought of as the 7-dimensional generalization of the 3-dimensional case
\begin{equation}
\tau^{i}\tau^{j} = \delta^{ij} + i e^{ijk}\tau^{k},
\end{equation}
where $\tau^{i}$ are the ordinary Pauli matrices.

\vspace{2mm}
Coming back to the equation \eqref{eq600} and applying the octonion identity we obtain
\begin{eqnarray}
(\Gamma^{ij})_{(+k)(-0)}\hat{\bar{D}}_{-k}\hat{\bar{D}}_{+0}C_{(+0)(+0)(+0)} &\propto & (\Gamma^{ij})_{(+0)}^{\hspace{6mm}E}C_{E(+0)(+0)}\nonumber\\
\epsilon^{ijk}\hat{\bar{D}}_{-k}\hat{\bar{D}}_{+0}C_{(+0)(+0)(+0)} &\propto & \epsilon^{ijk}C_{(+k)(+0)(+0)}
\end{eqnarray}
Therefore, we have obtained the state $C_{(+i)(+0)(+0)}$. By acting with $R^{-k}$ on $C_{(+0)(+0)(+0)}$ we obtain the state $C_{(-i)(+0)(+0)}$:
\begin{eqnarray}
(\Gamma^{-k})_{(+i)(+j)}\hat{\bar{D}}_{-i}\hat{\bar{D}}_{-j}C_{(+0)(+0)(+0)} &\propto & (\Gamma^{-k})_{(+0)}^{\hspace{6mm}E}C_{E(+0)(+0)}\nonumber\\
\epsilon^{kij}\hat{\bar{D}}_{-i}\hat{\bar{D}}_{-j}C_{(+0)(+0)(+0)} &\propto & \delta^{kl}C_{(-l)(+0)(+0)}
\end{eqnarray}

In this way, one can obtain all states contained in $C_{ABC}$. The table below shows explicitly how this is done. For brevity, we include only one way to obtain each state. The dash ($-$) means that all states corresponding to an initial state have been already obtained from other initial states. Finally, since $C_{ABC}$ is completely symmetric, states related by symmetry to states on the table need not be included.

\begin{table}
\caption{States produced by the rotation operator $R^{IJ}$}
\begin{center}
    \begin{tabular}{  | l | l | p{8cm} |}
    \hline
    Initial state & States produced by $R^{ij}$ & States produced by $R^{-k}$  \\ \hline
    $C_{(+0)(+0)(+0)}$ & $C_{(+k)(+0)(+0)}$ & $C_{(-k)(+0)(+0)}$  
     \\ \hline
    $C_{(+k)(+0)(+0)}$ &  $C_{(+k)(+l)(+0)}$ & $C_{(-0)(+0)(+0)}$, $C_{(+l)(-j)(+0)}$  \\ \hline
    $C_{(+k)(+l)(+0)}$ &  
         $C_{(+k)(+l)(+r)}$ & 
          $C_{(+k)(-0)(+0)}$, $C_{(+k)(+l)(-r)}$ \\
    \hline
    $C_{(-0)(+0)(+0)}$ &  
    - & $C_{(-0)(-k)(+0)}$ \\
    \hline
    $C_{(+l)(-k)(+0)}$ &  $C_{(-j)(-r)(+0)}$ & $C_{(+l)(-r)(-k)}$  \\
    \hline
    $C_{(+l)(+k)(+r)}$ &  - & $C_{(-0)(+l)(+r)}$  \\
    \hline
    $C_{(+k)(-0)(+0)}$ &  - & $C_{(-0)(-0)(+0)}$,  $C_{(+k)(-0)(-r)}$ \\
    \hline
    $C_{(-0)(-k)(+0)}$ &  - & $C_{(-0)(-k)(-r)}$ \\
    \hline
    $C_{(-k)(-r)(+0)}$ &  - & $C_{(-k)(-r)(-t)}$ \\
    \hline
    $C_{(-0)(+l)(+r)}$ &  - & $C_{(-0)(-0)(+r)}$\\
    \hline 
    $C_{(-0)(-0)(+0)}$ &  - & $C_{(-0)(-0)(-r)}$\\
    \hline 
    $C_{(-0)(-0)(+r)}$ &  - & $C_{(-0)(-0)(-0)}$\\
    \hline 
    
    \end{tabular}
    
\end{center}
\end{table}
\end{appendices}

\newpage
\bibliographystyle{utphys}
\providecommand{\href}[2]{#2}\begingroup\raggedright\endgroup

\end{document}